\documentclass[aps,nofootinbib,prd,twocolumn]{revtex4-2}

\usepackage[english]{babel}
\usepackage[utf8]{inputenc}
\usepackage{amsthm}
\usepackage{mathtools}
\usepackage{physics}
\usepackage{xcolor}
\usepackage{graphicx}
\usepackage[left=23mm,right=13mm,top=35mm,columnsep=15pt]{geometry} 
\usepackage{adjustbox}
\usepackage{placeins}
\usepackage[T1]{fontenc}
\usepackage{csquotes}
\usepackage[pdftex, pdftitle={Article}, pdfauthor={Author}]{hyperref} 
\usepackage{setspace}
\usepackage{float}
\usepackage{graphicx}
\usepackage{subcaption}
\usepackage{booktabs}  
\usepackage{url}
\usepackage{hyperref}

\usepackage{array}
\usepackage{orcidlink}

\newcommand{\ab}{ab\ensuremath{^{-1}}}
\newcommand{\sig}{\ensuremath{\sigma}}
\newcommand{\ZZ}{\ensuremath{ZZ}}
\newcommand{\W}{\ensuremath{WW}}
\newcommand{\ttbar}{$t\bar{t}$}

\newcommand{\chan}[2]{\texttt{Z#1H#2}}

\usepackage{booktabs}
\usepackage{multirow}

\begin{document}

\preprint{}

\title{Measurements of $H\rightarrow W^+W^-$ in the Fully Leptonic Decay Mode at the FCC-ee}

\author{Kael Kemp \orcidlink{0009-0002-8587-6425}}%
 \email{kael.kemp@adelaide.edu.au}
\affiliation{Department of Physics, Adelaide University, North Terrace, Adelaide, SA 5005, Australia}

\author{Aman Desai \orcidlink{0000-0003-2631-9696}}
\email{aman.desai@adelaide.edu.au}
\affiliation{Department of Physics, Adelaide University, North Terrace, Adelaide, SA 5005, Australia}

\author{Paul Jackson \orcidlink{0000-0002-0847-402X}}
\email{p.jackson@adelaide.edu.au}
\affiliation{Department of Physics, Adelaide University, North Terrace, Adelaide, SA 5005, Australia}

\date{\today}
\onehalfspacing
\renewcommand{\arraystretch}{1.45} 
\begin{abstract}

The expected precision on measuring the $\sigma(e^+ e^- \rightarrow ZH) \times Br(H\rightarrow W^+W^-)$ in the fully leptonic decay mode at the Future Circular Collider (FCC) is presented. We consider two FCC-ee scenarios: $\sqrt{s} =240$ GeV centre-of-mass energy with a luminosity of 10.8 \ab~and $\sqrt{s} =365$ GeV centre-of-mass energy with a luminosity of 3.12 \ab. Our results indicate that a relative uncertainty of 2.9\% and 6.8\% can be achieved on measurements of $\sigma(e^+ e^- \rightarrow ZH) \times Br(H\rightarrow W^+W^-)$ in the fully leptonic decay mode at $\sqrt{s} =240$ GeV and $\sqrt{s} =365$ GeV, respectively. 

\end{abstract}
\maketitle


\section{Introduction}

The discovery of Higgs boson by the ATLAS and CMS collaborations at the Large Hadron Collider (LHC) completed the Standard Model~\cite{HIGG-2012-27,CMS-HIG-12-028,Englert:1964et,Higgs:1964ia}. The Higgs boson is the key to understanding the electroweak symmetry breaking mechanism, which gives masses to the $Z$ and $W^\pm$ gauge bosons. The Higgs boson couplings to the gauge bosons and third generation fermions have been studied at the LHC, showing consistency with the Standard Model predictions~\cite{HIGG-2021-23, CMS-HIG-22-001}.  Studying the properties of the Higgs boson with precision is one of the major goals of the Future Circular Collider (FCC), a proposed collider facility at CERN~\cite{FCC:2025lpp,FCC:2025jtd,FCC:2025uan}. The FCC-ee will collide electrons and positrons at centre-of-mass energies ranging from the $Z$-pole (91 GeV) up to the top-quark pair production threshold (365 GeV). Collisions at $\sqrt{s} = 240$ GeV will be important for performing Higgs measurements, as the Higgs production cross-section in the Higgs-strahlung ($ZH$) channel is maximal at this energy, $\sigma (e^+ e^- \rightarrow ZH) \approx  200\rm{~fb}$~\cite{FCC:2025lpp}. With a predicted integrated luminosity of $10.8\rm{~ab}^{-1}$ at this collision energy, FCC-ee would produce approximately $2.2 \times 10^6$ Higgs bosons. Additionally, Higgs measurements will also be possible at  $\sqrt{s} = 365$ GeV with a proposed integrated luminosity of $3.1\rm{~ab}^{-1}$, where the cross-section in the $ZH$ channel is  $\approx  125\rm{~fb}$~\cite{FCC:2025lpp}.

The branching fraction for the Higgs decay to $WW$\footnote{Throughout this paper, we assume charge conjugation.} is $\sim 21\%$, which makes it the second most common decay channel after the $b\bar{b}$ channel~\cite{ParticleDataGroup:2024cfk}. The Higgs coupling to $W$ bosons can be measured by analysing the $H\rightarrow WW$ decay channel. In this work, the decays of the $Z$ and $W$ bosons to leptons are considered, which gives rise to a final state consisting of four leptons and missing energy. The fully leptonic decay mode have relatively low branching fractions when compared to modes involving jets. The absence of pileup at FCC-ee and the final state consisting of no jets initiated by hadronisation of quarks leads to a higher precision measurement of $ZH[WW]$.

In this analysis, we study $Z(\ell\ell)H[WW(\ell\ell\nu\nu)]$, with $\ell$ being either an electron or a muon, at centre-of-mass energies $\sqrt{s} = 240, 365$ GeV. A survey of different decay modes - or - combination of electrons and muons is carried out. In particular, we consider the channels consisting of two electrons and two muons, three electrons (one electron) and one muon (three muons), four electrons, and four muons. We assess the impact of different leptonic decays on the relative uncertainty of the measurements.

This paper is organised as follows: in Section \ref{sec:event_sim}, we discuss the Monte Carlo samples used in this analysis. The preselection criteria used to prepare samples for the six different channels are discussed in Section \ref{sec:preselect}. In Section \ref{sec:strategy}, we present the analysis strategy, that includes a cut-flow and also a Machine Learning based approach. Finally, the signal strength when measuring the $\sigma(e^+ e^- \rightarrow ZH) \times Br(H\rightarrow WW)$ is determined in Section \ref{sec:sigstrength}.

\section{Event Simulation}\label{sec:event_sim}

\begin{table*}[htbp]
    \centering
    \caption{The list of processes considered in this analysis, their production cross-section at $\sqrt{s}$ = 240~GeV and number of events.}
    
        \label{tab:cross-section}
\small
\begin{tabular}{l l r r}
\toprule
Label & Process & Cross section [fb] & $N_\text{events}$ \\
\hline\hline

\multicolumn{4}{c}{\textbf{Signal}} \\
\midrule

\multirow{2}{*}{$ZH[WW]$} 
 & $\mathrm{e^+ e^-}\!\rightarrow Z(\mu^+\mu^-)H(W W^\ast)$ & 1.46 & 400,000 \\
 & $\mathrm{e^+ e^-}\!\rightarrow Z(e^+e^-)H(W W^\ast)$     & 1.54 & 400,000 \\

\midrule[\heavyrulewidth]

\multicolumn{4}{c}{\textbf{Background}} \\
\midrule

$WW$ 
 & $\mathrm{e^+ e^-}\!\rightarrow W^+W^-$ & $1.64\times10^{4}$ & $3.73\times10^8$ \\
$ZZ$ 
 & $\mathrm{e^+ e^-}\!\rightarrow Z Z$    & $1.36\times10^{3}$ & $5.62\times10^8$ \\

\midrule

\multirow{7}{*}{$ZH$}
 & $\mathrm{e^+ e^-}\!\rightarrow Z(\tau^+\tau^-)H(W W^\ast)$   & 1.45                & 400,000 \\
 & $\mathrm{e^+ e^-}\!\rightarrow Z(\mu^+\mu^-)H(Z Z^\ast)$     & 0.18  & 400,000 \\
 & $\mathrm{e^+ e^-}\!\rightarrow Z(e^+e^-)H(Z Z^\ast)$         & 0.19 & 400,000 \\
 & $\mathrm{e^+ e^-}\!\rightarrow Z(\nu\bar{\nu})H(Z Z^\ast)$   & 1.22                & 1,200,000 \\
 & $\mathrm{e^+ e^-}\!\rightarrow Z(e^+e^-)H(\tau^+\tau^-)$     & 4.49                & 400,000 \\
 & $\mathrm{e^+ e^-}\!\rightarrow Z(\mu^+\mu^-)H(\tau^+\tau^-)$ & 0.42                & 400,000 \\
 & $\mathrm{e^+ e^-}\!\rightarrow Z(\tau^+\tau^-)H(\mu^+\mu^-)$ & $1.47\times10^{-3}$ & 400,000 \\
 & $\mathrm{e^+ e^-}\!\rightarrow Z(\tau^+\tau^-)H(ZZ^\ast)$    & 0.18 & 330,996\\
 & $\mathrm{e^+ e^-}\!\rightarrow Z(\tau^+\tau^-)H(\tau^+\tau^-)$& 0.42&  400,000 \\
\bottomrule
\end{tabular}

    \end{table*}

\begin{table*}[tbp]
    \centering
    \caption{The list of processes considered in this analysis, their production cross-section at $\sqrt{s}$ = 365 GeV and number of events.}
        \label{tab:cross-section365}

\small

\begin{tabular}{l l r r}
\toprule
Label & Process & Cross section [fb] & $N_\text{events}$ \\
\hline\hline

\multicolumn{4}{c}{\textbf{Signal}} \\
\midrule

\multirow{2}{*}{$ZH[WW]$} 
 & $\mathrm{e^+ e^-}\!\rightarrow Z(\mu^+\mu^-)H(W W^\ast)$ & 0.90 &  1,100,000 \\
 & $\mathrm{e^+ e^-}\!\rightarrow e^+e^-H(W W^\ast)$     & 1.59 & 1,100,000 \\

\midrule[\heavyrulewidth]

\multicolumn{4}{c}{\textbf{Background}} \\
\midrule

$WW$ 
 & $\mathrm{e^+ e^-}\!\rightarrow W^+W^-$ & $1.07\times10^4$ & $1.01\times10^8$\\
$ZZ$ 
 & $\mathrm{e^+ e^-}\!\rightarrow Z Z$    & $6.43\times10^2$ & $6.14 \times 10^{7}$ \\
 $t\bar{t}$
 & $\mathrm{e^+ e^-}\!\rightarrow t \bar{t}$ & $8.00\times10^2$& 2,700,000 \\

\midrule

\multirow{7}{*}{ZH}
 & $\mathrm{e^+ e^-}\!\rightarrow Z(\tau^+\tau^-)H(W W^\ast)$ & 0.90 &  1,100,000 \\
 & $\mathrm{e^+ e^-}\!\rightarrow Z(\mu^+\mu^-)H(Z Z^\ast)$   & 0.11 & 800,000 \\
 & $\mathrm{e^+ e^-}\!\rightarrow e^+e^-H(Z Z^\ast)$       & 0.20 & 1,200,000 \\
 & $\mathrm{e^+ e^-}\!\rightarrow \nu\bar{\nu}H(Z Z^\ast)$ & 1.42 & 1,200,000 \\
 & $\mathrm{e^+ e^-}\!\rightarrow e^+e^-H(\tau^+\tau^-)$   & 0.46 & 1,200,000 \\
 & $\mathrm{e^+ e^-}\!\rightarrow Z(\mu^+\mu^-)H(\tau^+\tau^-)$ & 0.26 & 900,000 \\
 & $\mathrm{e^+ e^-}\!\rightarrow Z(\tau^+\tau^-)H(\mu^+\mu^-)$ & $9.08\times10^{-4}$ & 1,200,000 \\
 & $\mathrm{e^+ e^-}\!\rightarrow Z(\tau^+\tau^-)H(ZZ^\ast)$ & 1.10 & 1,200,000 \\
 & $\mathrm{e^+ e^-}\!\rightarrow Z(\tau^+\tau^-)H(\tau^+\tau^-)$& 0.26 & 1,100,000\\
\bottomrule
\end{tabular}
\end{table*}


The Monte Carlo samples used in this study were produced centrally by the \textsc{FCC} Collaboration and are tagged as the \textsc{Winter2023} campaign~\cite{FCC_EE_IDEA_Winter2023}. The signal and the background samples used in this analysis that included a Higgs boson were generated using \textsc{Whizard}~\cite{Kilian:2007gr} followed by \textsc{Pythia6}~\cite{Sjostrand:2006za} for hadronisation and parton shower. The decays of $H\rightarrow WW$ and $H\rightarrow ZZ$ are inclusive. The background samples, \ZZ, \W, and \ttbar, were generated using \textsc{Pythia8}~\cite{Sjostrand:2014zea} and also consider inclusive decays of the gauge bosons. 

The generated events are then passed through the \textsc{Delphes} fast simulation package~\cite{deFavereau:2013fsa} to simulate the parametric response of the proposed \textsc{IDEA} detector~\cite{IDEAStudyGroup:2025gbt}. The implementation is such that electron and muon tracks with $p_T>100~\rm{MeV}$  and $|\eta| < 2.56$ are assumed to be reconstructed with 100\% efficiency. Smearing is applied, taking into account the detector resolution and also scattering of particles by detector material. Identification efficiency for electrons and muons is set to 99\% for $E>2~\rm{GeV}$  and $|\eta| < 3$. The imbalance in energy is identified as $E^{\rm miss}$. 

This event generation chain is implemented in the \textsc{Key4HEP} software~\cite{Key4hep:2023nmr}. The configuration files used in the production can be found in Ref.~\cite{HEP-FCC_FCC-config}. 

In this study, the right-handed coordinate system is used. The interaction point is the centre of the IDEA detector and is used as the origin of this coordinate system. The x-axis points towards the centre of the FCC-ee and the y-axis points upwards. The z-axis points in the beam direction, and the azimuthal angle is measured from the x-axis.

\autoref{tab:cross-section} gives a list of all the samples used in this study for $\sqrt{s} = 240$~GeV, along with the cross-sections, and the number of events generated. The corresponding table for $\sqrt{s} = 365~\rm{GeV}$ is given in \autoref{tab:cross-section365}. At this energy, contribution from Vector Boson Fusion is significant in processes such as $e^+e^-H$ and $\nu\bar{\nu} H$. The samples labelled as $ZH[WW]$ consider the decay of the $Z$ boson to either a pair of electrons or a pair of muons.

\section{Event Preselection}\label{sec:preselect}

We carry out these analyses using the \textsc{FCCAnalyses} software framework~\cite{helsens_2025_15528870}. The analysis for $\sqrt{s} = 240~\rm{GeV}$ and $\sqrt{s} = 365~\rm{GeV}$ is carried over six orthogonal final states which are listed in \autoref{tab:channels}. 

\begin{table}[htbp]
\centering
\caption{The leptonic decay modes considereded in this study and their corresponding labels.}
\label{tab:channels}
\begin{tabular}{c|c}
\hline
\textbf{Final State} & \textbf{Label} \\
\hline\hline
$Z(\mu\mu),H[WW(ee\nu\nu)]$       & \chan{mumu}{ee}   \\
$Z(\mu\mu),H[WW(e\mu\nu\nu)]$    & \chan{mumu}{emu}  \\
$Z(\mu\mu),H[WW(\mu\mu\nu\nu)]$  & \chan{mumu}{mumu} \\
$Z(ee),H[WW(ee\nu\nu)]$          & \chan{ee}{ee}     \\
$Z(ee),H[WW(e\mu\nu\nu)]$        & \chan{ee}{emu}    \\
$Z(ee),H[WW(\mu\mu\nu\nu)]$      & \chan{ee}{mumu}   \\
\hline
\end{tabular}
\end{table}

As a general strategy, events consisting of at least four leptons satisfying the condition $|\Vec{p_{\ell}}| > 5$ GeV  are preselected. As our final state consists of neutrinos, which escape the detector without detection, we also require $E_{\rm miss} > 5$ GeV. 

Each final state considered in this study is processed through a set of channel specific preselection criteria. These criteria are in place to ensure the different sub-analyses remain orthogonal -- which is useful when carrying out statistical combination of different channels.  We achieve this in the following way: in the \chan{ee}{mumu} and \chan{mumu}{ee} channels we require exactly two electrons and two muons, $N_e = 2$ and $N_\mu = 2$ with an additional requirement of net charge of the muons and electrons being zero: $\sum_{e} N_{Q} = 0$ and $\sum_{\mu} N_{Q} = 0$.  For the \chan{ee}{emu} (\chan{mumu}{emu}) channels we require $N_e = 3$, $N_\mu = 1$,  ($N_\mu = 3$ and $N_e = 1$). For these channels, we require that the net charge of all leptons to be equal to zero. For the \chan{ee}{ee} (\chan{mumu}{mumu}), we require $N_e = 4$ and $N_\mu = 0$ ($N_\mu = 4$ and $N_e = 0$). Again, the net charge of all leptons is required to be equal to zero.

The final state leptons are then used to reconstruct the $Z$ boson, considering the $Z$ boson having a mass of 91~GeV. In \chan{ee}{mumu} (\chan{mumu}{ee}), which consists of exactly two electrons (muons) coming from the decay of the $Z$ boson, the $Z$ boson is reconstructed by summing the four-momenta of the same flavour but opposite sign leptons. In the \chan{ee}{emu} and \chan{mumu}{emu} channels, an additional combinatoric step is performed to find the combination of leptons with net-zero charge, the same flavour and their invariant mass closest to the $Z$~boson mass. The \chan{ee}{ee} and \chan{mumu}{mumu} present a combinatorics challenge, wherein the best resonance is to be found from a set of combinations of electrons or muons, with opposite sign, and choosing the pair which gives the least difference in the invariant mass of lepton pair and the $Z$ boson mass. 

The pair of leptons that is not used to reconstruct the $Z$ boson resonance is then assumed to be coming from the $H\rightarrow WW$ decays. We label this pair separately, as these variables can enable us to discriminate against the background arising from $ZZ$. 

Furthermore, we also reconstruct the recoil mass as the $Z$ boson recoils against the Higgs boson. This variable is computed based on energy-momentum conservation and also information about the initial state, which is well known at the $e^+e^-$ collider. This quantity in the $Z(\ell\ell)H$ case is given as: 

\begin{equation}
    m_{\rm recoil}^{2} = s + m_{Z}^2 - 2 E_Z \sqrt{s}
\end{equation}

where the energy and the invariant mass of the $Z$ boson are computed by combining the lepton pair as discussed earlier. The variable is such that it peaks near the invariant mass of the recoiling object, which in the case of the signal, is the Higgs boson mass (125 GeV).  As the background $ZZ$ and $WW$ do not contain a Higgs boson, they are not likely to peak near the Higgs mass, allowing us to use this quantity as a handle to discriminate signal from the background.

The resulting yield after applying these preselection conditions are summarised for $\sqrt{s} = 240~\rm{GeV}$ and $\sqrt{s} = 365~\rm{GeV}$ in \autoref{tab:cutFlow240} and \autoref{tab:cutFlow365}, respectively. The $ZZ$ process remains a dominant source of background across all channels.

\begin{figure*}[tbp]
        \subfloat[\centering \chan{ee}{emu}]{\includegraphics[width=0.32\linewidth]{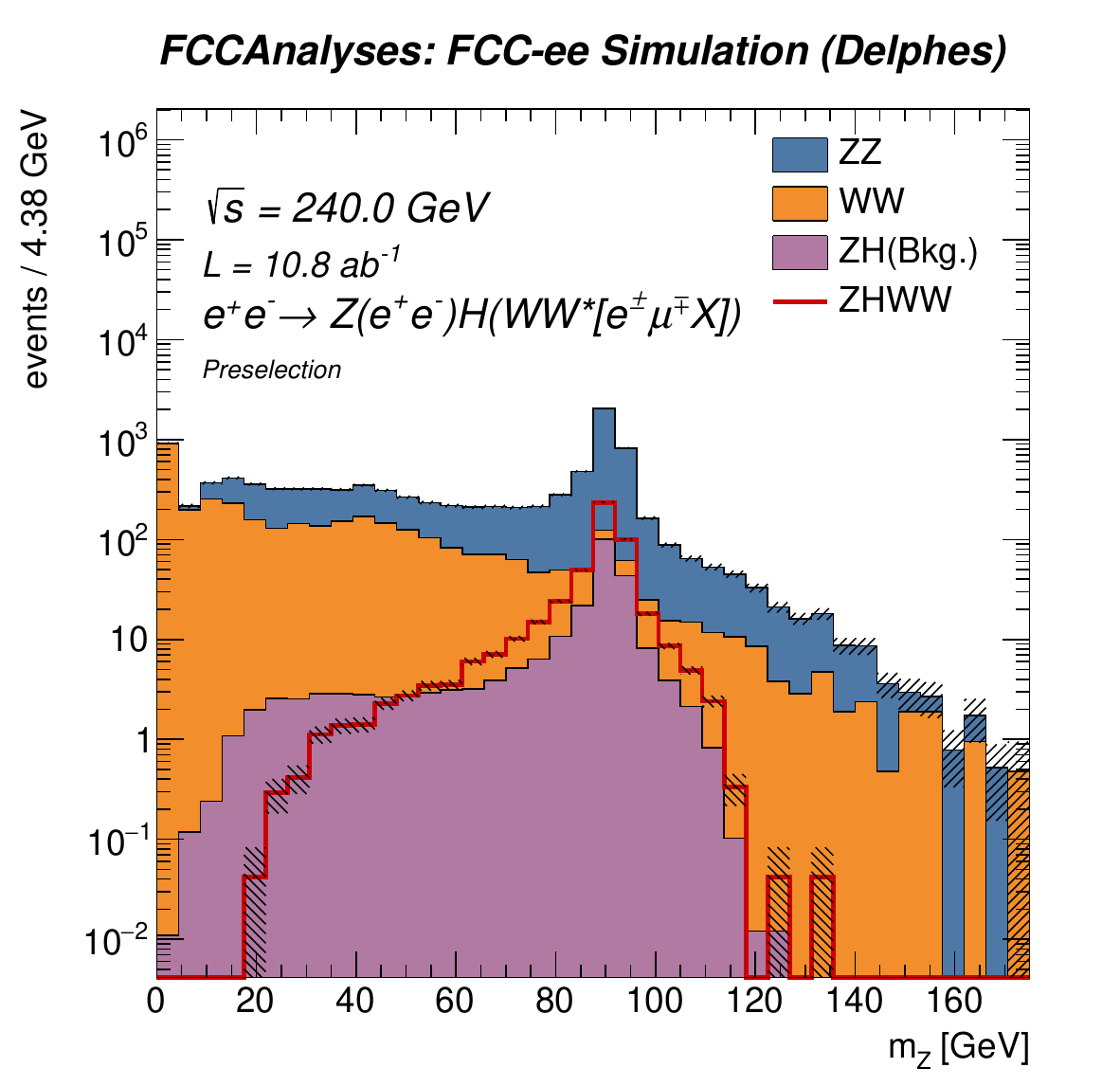}}
        \subfloat[\centering \chan{mumu}{emu}]{\includegraphics[width=0.32\linewidth]{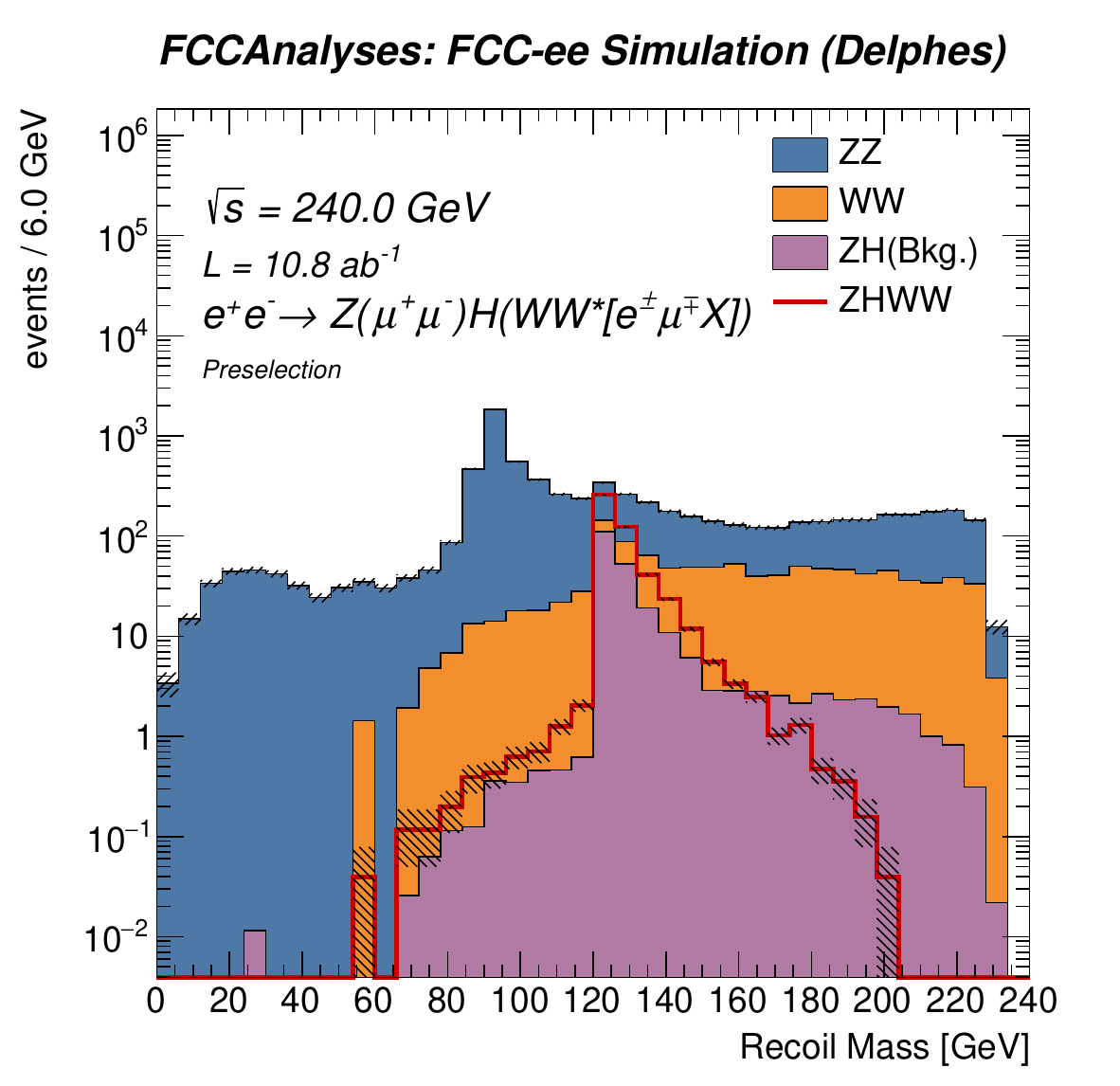}}
        \subfloat[\centering \chan{ee}{mumu}]{\includegraphics[width=0.32\linewidth]{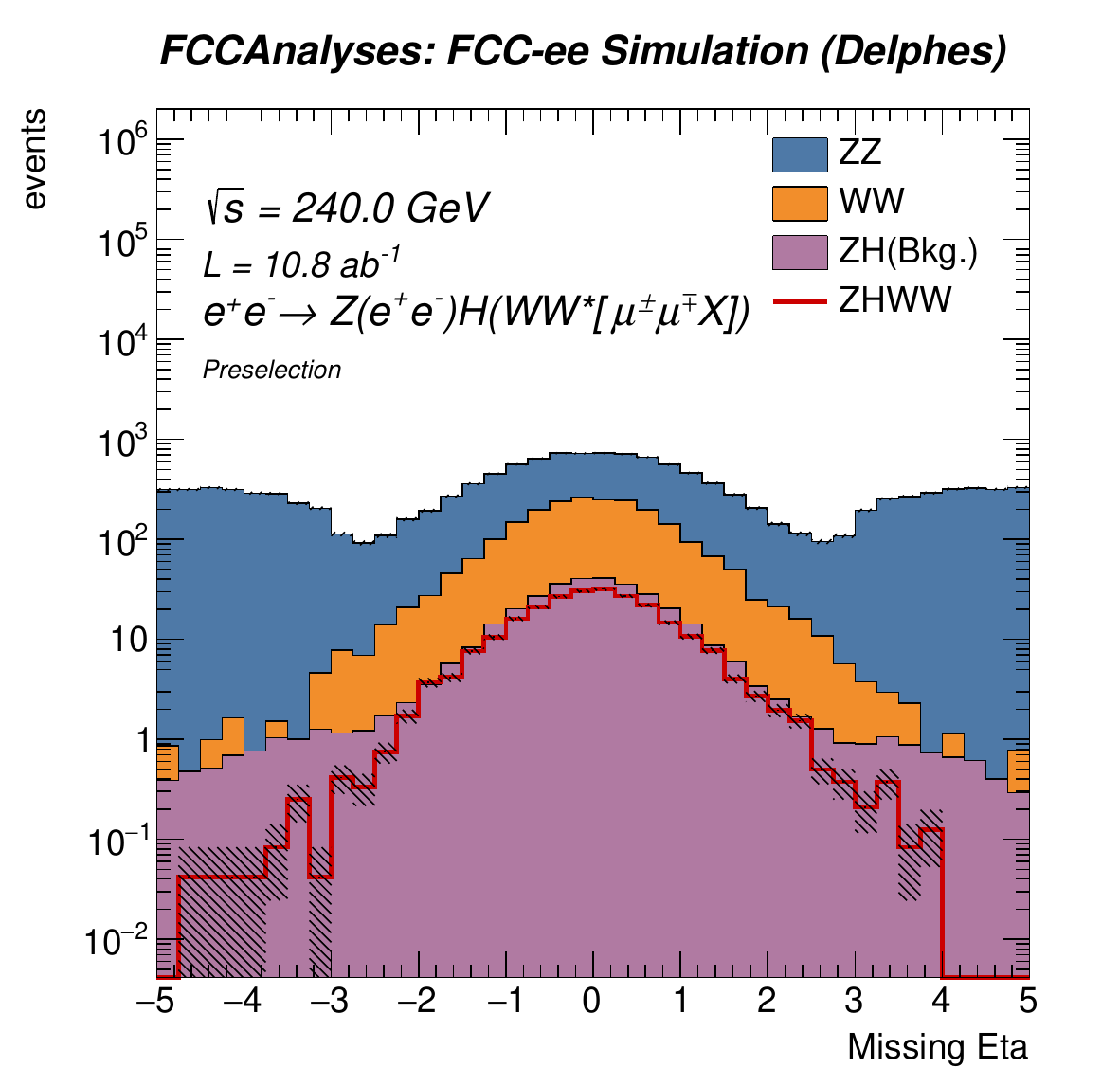}}
        \qquad
        \subfloat[\centering \chan{ee}{ee}]{\includegraphics[width=0.32\linewidth]{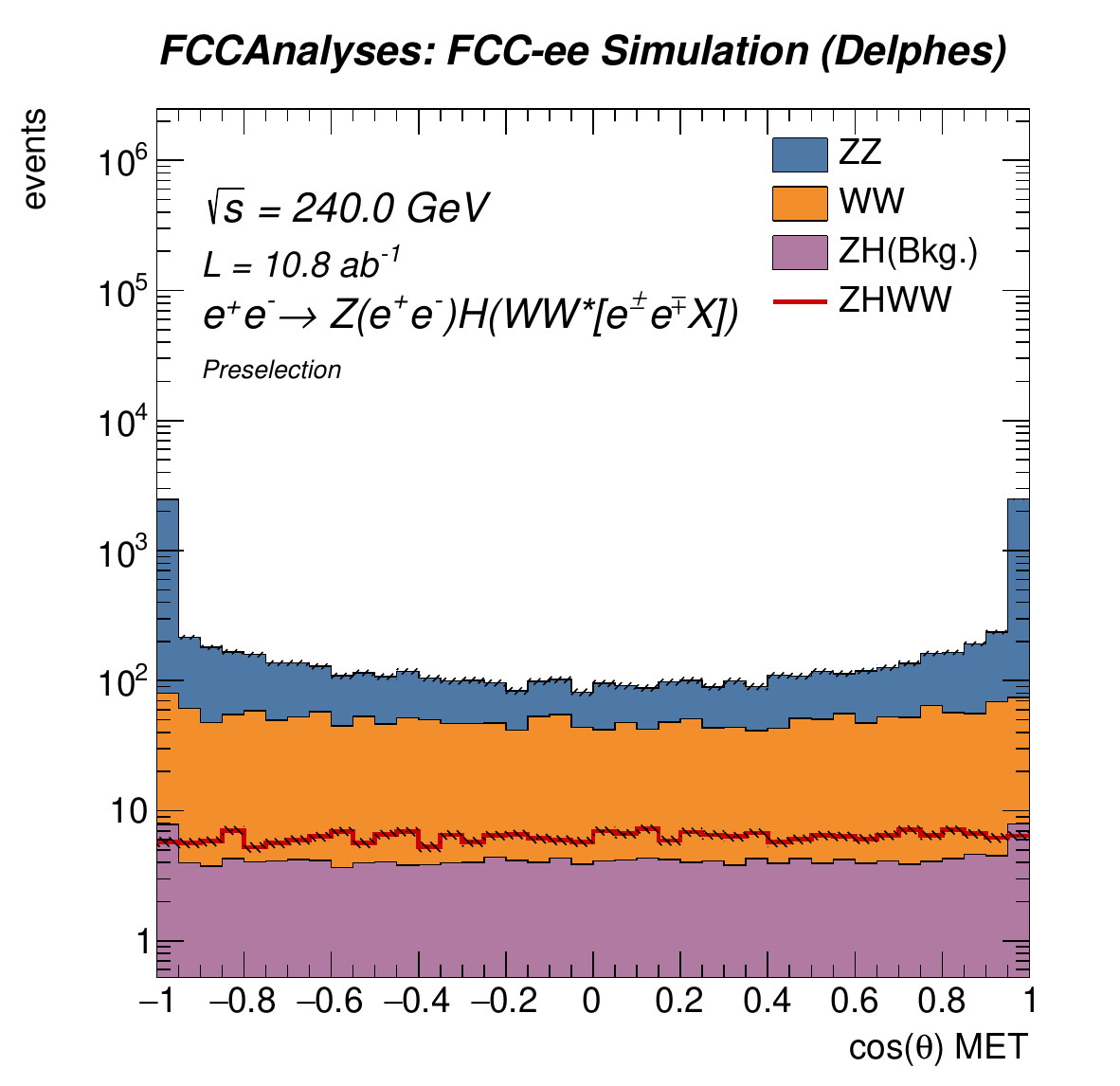}}
        \subfloat[\centering \chan{mumu}{ee}]{\includegraphics[width=0.32\linewidth]{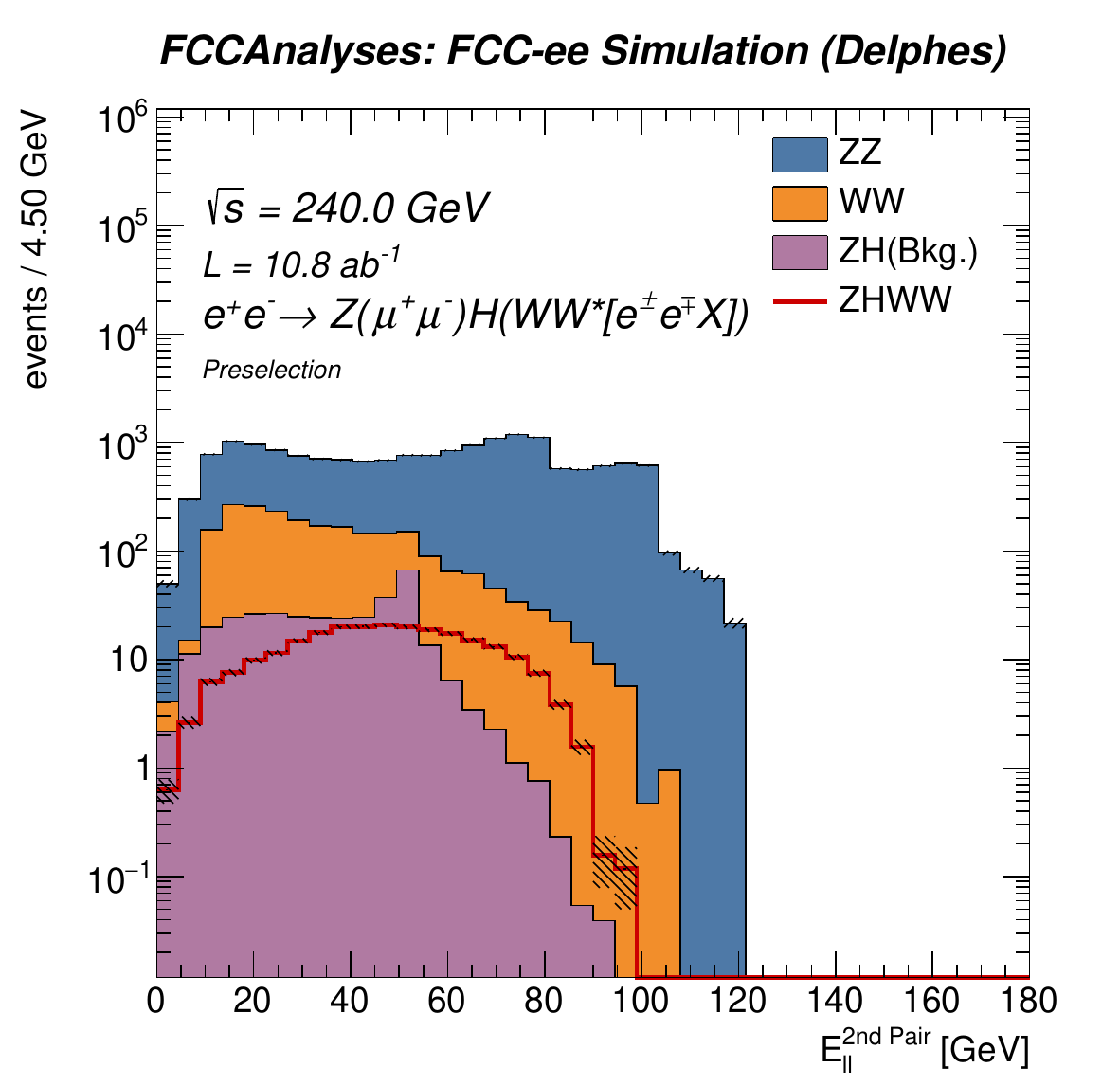}}
        \subfloat[\centering \chan{mumu}{mumu}]{\includegraphics[width=0.32\linewidth]{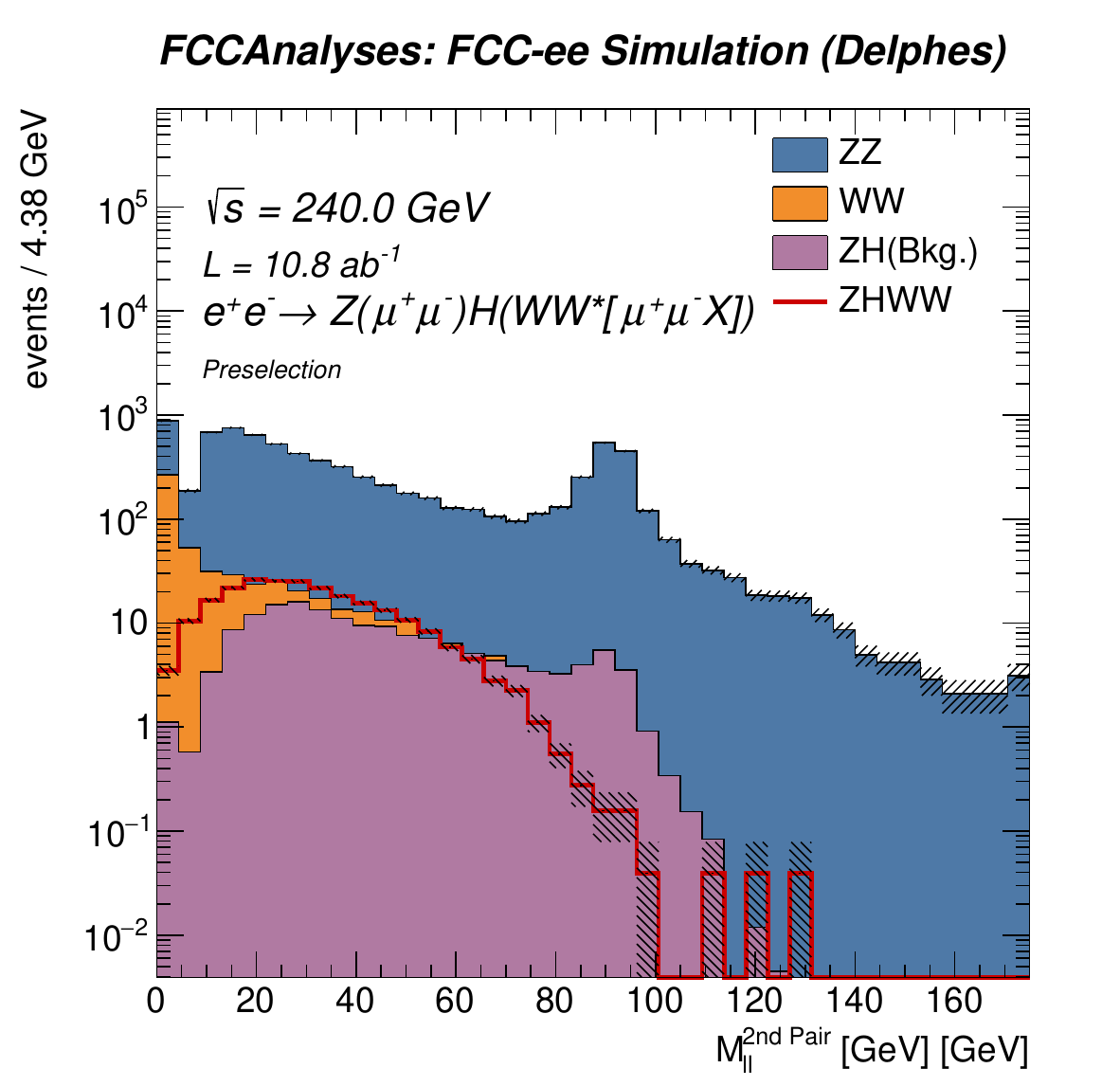}}
        \caption{Distributions of observables after applying preselection criteria for $\sqrt{s}=240$ GeV. (a) $Z$ candidate mass for the \chan{ee}{emu} channel. (b) Recoil mass for the \chan{mumu}{emu} channel. (c) $\eta_{\text{miss}}$ for the \chan{ee}{mumu} channel. (d)  $\cos{\theta_\text{MET}}$ in the \chan{ee}{ee} channel. (e)  energy of the second lepton pair in the \chan{mumu}{ee} channel. (f) invariant mass of the second lepton pair in the \chan{mumu}{mumu} channel. The uncertainty associated with the MC simulation is shown by the shaded area.}%
        \label{fig:Variables240}
\end{figure*}

\begin{figure*}[tbp]
        \subfloat[\centering \chan{ee}{emu}]{\includegraphics[width=0.32\linewidth]{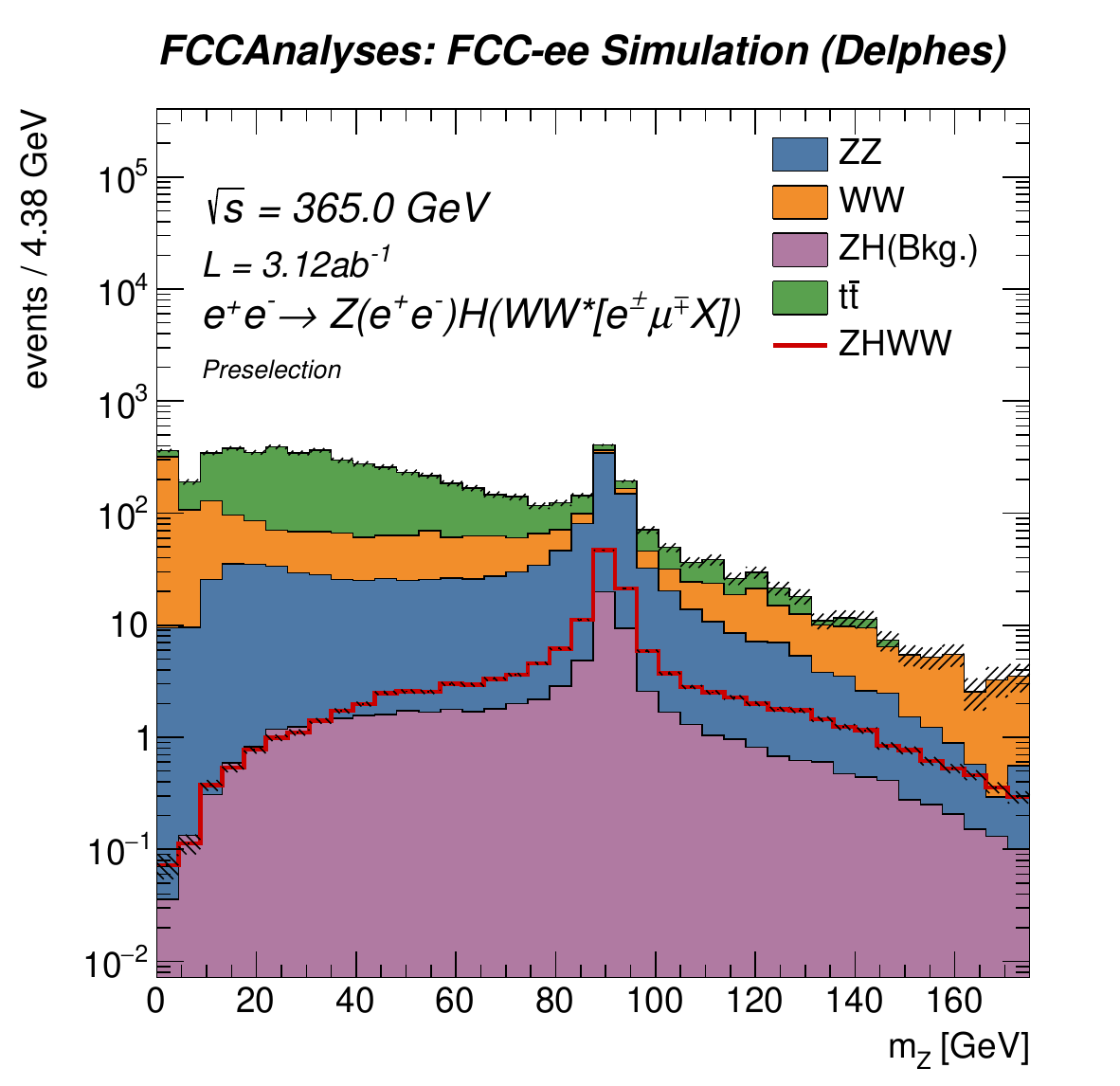}}
        \subfloat[\centering \chan{mumu}{emu}]{\includegraphics[width=0.32\linewidth]{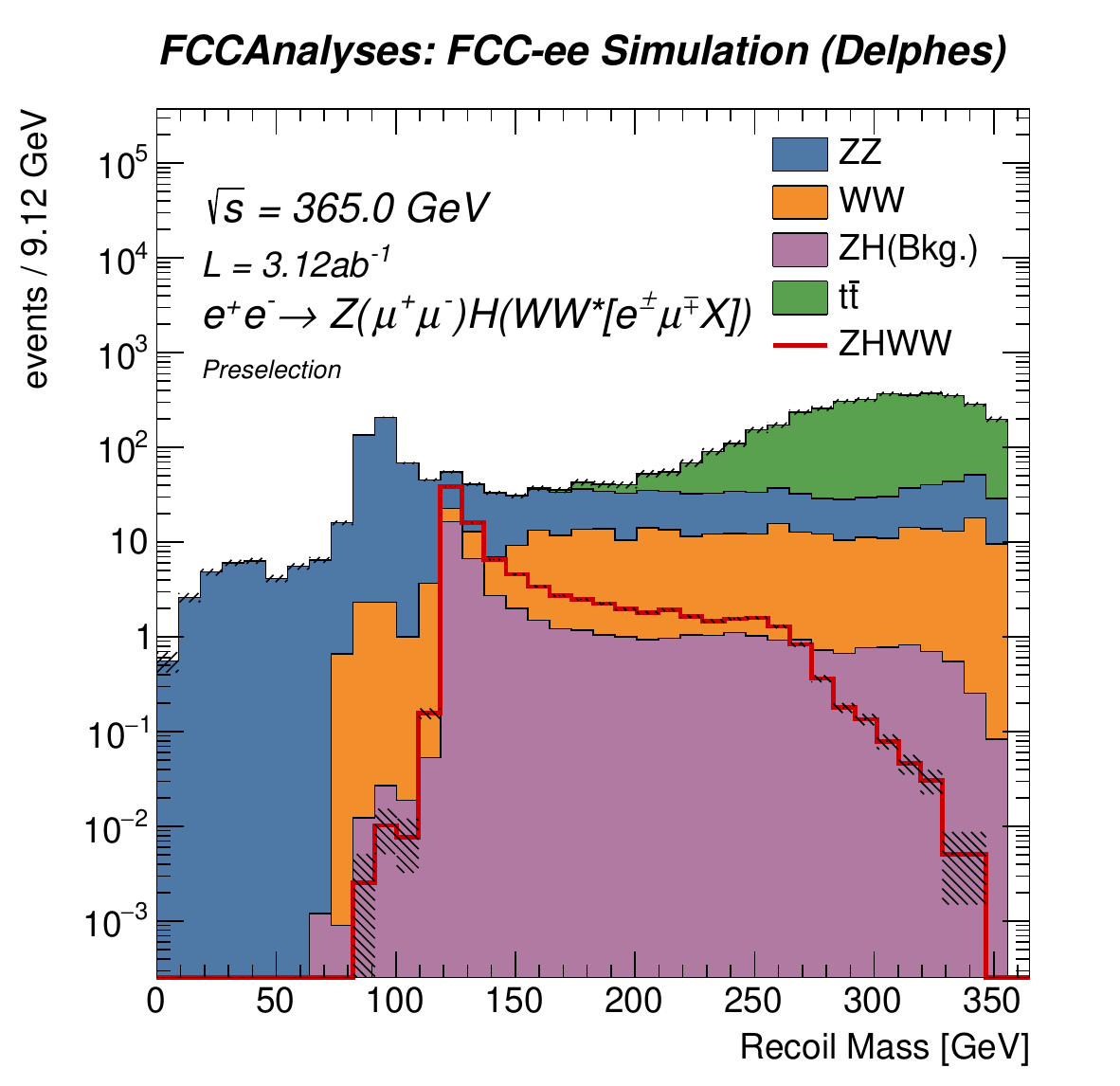}}
        \subfloat[\centering \chan{ee}{mumu}]{\includegraphics[width=0.32\linewidth]{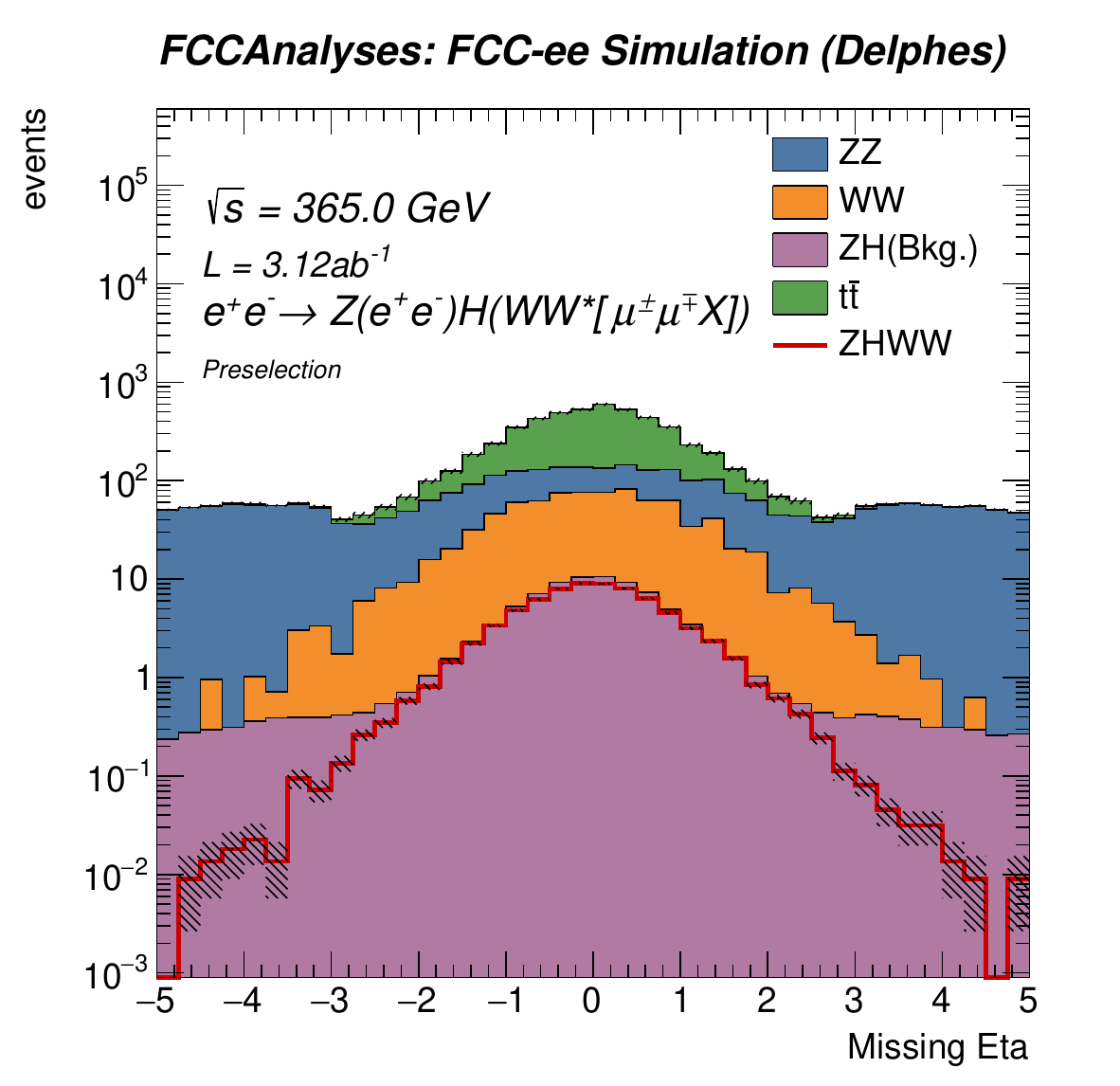}}
        \qquad
        \subfloat[\centering \chan{ee}{ee}]{\includegraphics[width=0.32\linewidth]{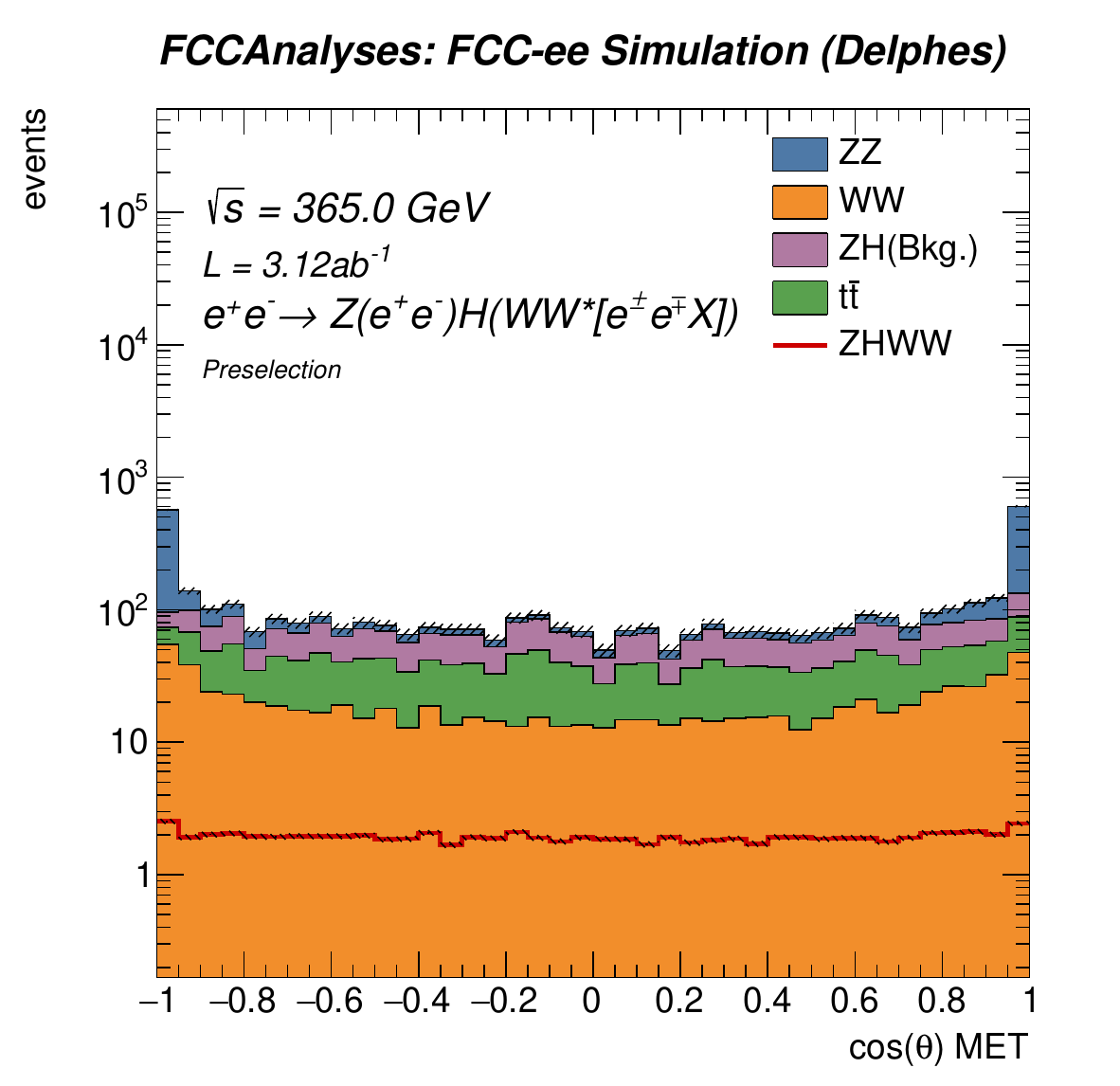}}
        \subfloat[\centering \chan{mumu}{ee}]{\includegraphics[width=0.32\linewidth]{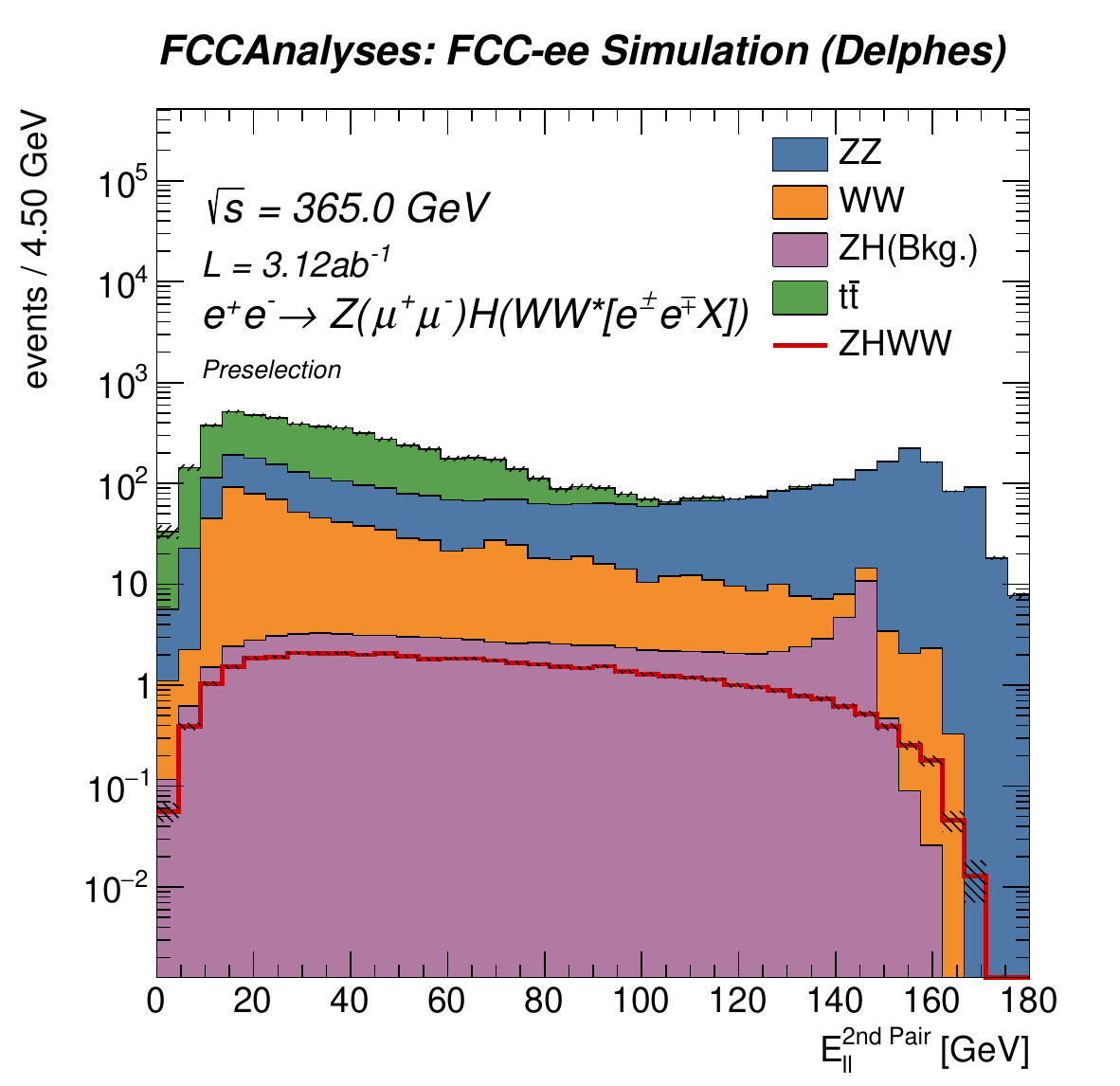}}
        \subfloat[\centering \chan{mumu}{mumu}]{\includegraphics[width=0.32\linewidth]{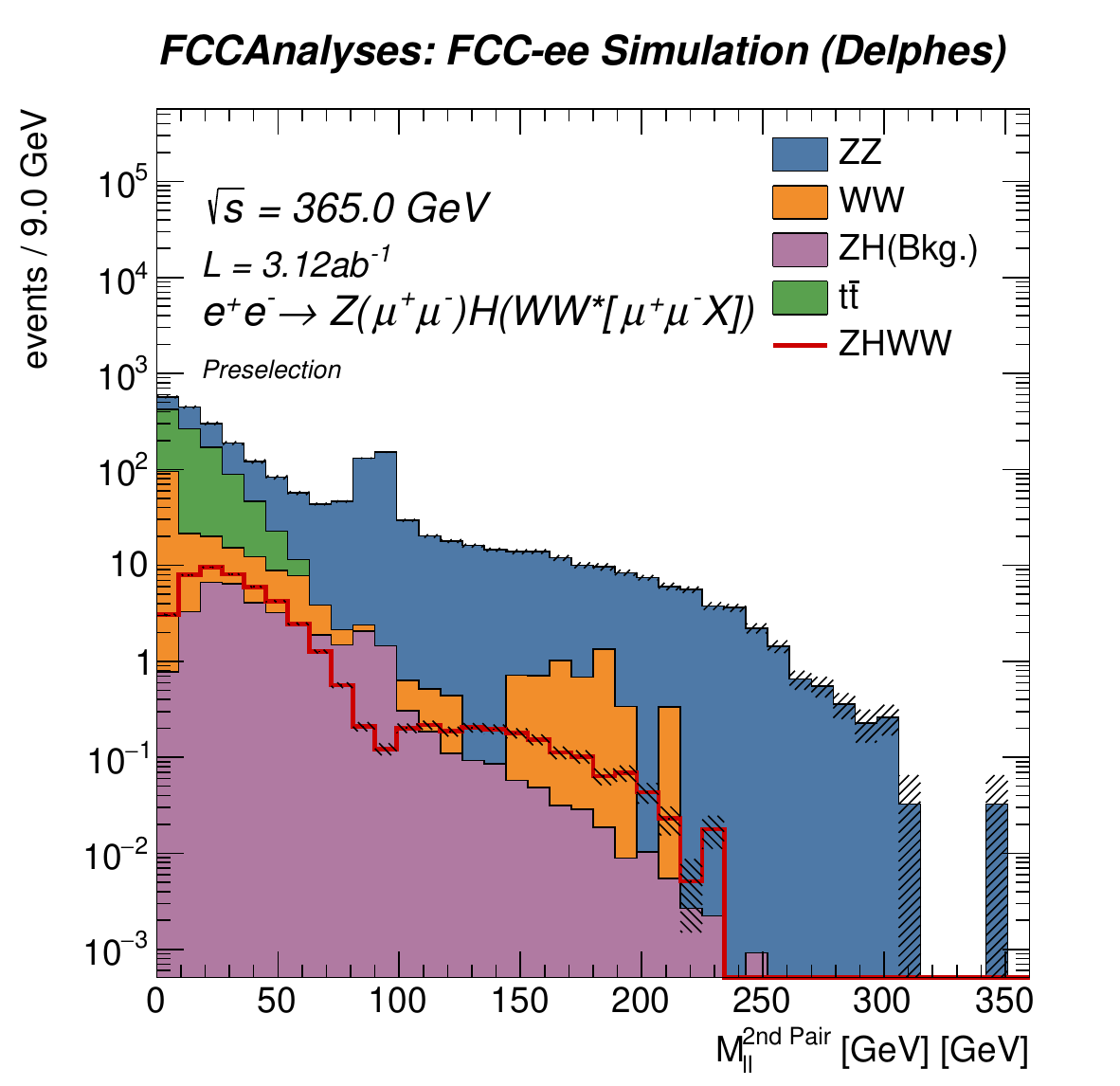}}
        \caption{Distributions of observables after applying preselection criteria for $\sqrt{s}=365$ GeV. (a) $Z$ candidate mass for the \chan{ee}{emu} channel. (b) Recoil mass for the \chan{mumu}{emu} channel. (c) $\eta_{\text{miss}}$ for the \chan{ee}{mumu} channel. (d)  $\cos{\theta_\text{MET}}$ in the \chan{ee}{ee} channel. (e)  energy of the second lepton pair in the \chan{mumu}{ee} channel. (f) invariant mass of the second lepton pair in the \chan{mumu}{mumu} channel. The uncertainty associated with the MC simulation is shown by the shaded area.}%
        \label{fig:Variables365}
\end{figure*}

The distributions of the variables studied in this analysis, after applying preselection criteria, are given in \autoref{fig:Variables240} ($\sqrt{s} = 240~\rm{GeV}$) and \autoref{fig:Variables365} ($\sqrt{s} = 365~\rm{GeV}$).

\section{Analysis Strategy}\label{sec:strategy}

For each final state the analysis is performed separately in two stages. The first stage involves determining a set of selection criterion based on the distributions of the observables, motivated by improving significance at each step by rejecting the background events. Here, the significance is defined as $S/\sqrt{S+B}$. In the second stage, Multivariate Analysis (MVA) methods are applied to the signal and background events to find a discriminator that could improve the significance of the observation. 

In \autoref{tab:cutFlow240}, we present a set of selection criteria and the corresponding yields for the signal and background processes at $\sqrt{s}=240$ GeV. The preselection criteria defined earlier lead to an initial significance of $\approx 1.8\sigma$ in the \chan{mumu}{ee} and $\approx 1.9\sigma$ in the \chan{ee}{mumu} channels. In the channels consisting of four electrons or four muons, the preselection significance is $\approx 2.5\sigma$ and $2.6\sigma$, respectively.  Due to the absence of major backgrounds consisting of three muons and one electron  (one muon and three electrons), the \chan{mumu}{emu} (\chan{ee}{emu}) show a significance of about $\approx 5.4~ (4.9)\sigma$.

\begin{table*}[htbp]
\centering
\caption{Event yields and significance for all channels at $\sqrt{s}=240$ GeV after applying selection conditions.}
\label{tab:cutFlow240}
\begin{tabular}{cc}

\begin{subtable}{0.48\textwidth}
\centering
\caption{\chan{mumu}{ee}}
\resizebox{\textwidth}{!}{
\begin{tabular}{l r r r r r r}
\toprule
Selection & $S$ & $B_{\mathrm{ZZ}}$ & $B_{\mathrm{WW}}$ & $B_{ZH}$ & $B_{\rm tot}$ & $Z$ \\
\hline
\hline

Preselection & 240 & 15139 & 1947 & 356 & 17442 & 1.8 \\
$80<m_Z<100$ GeV & 216 & 6821 & 171 & 161 & 7152 & 2.5 \\
$p_{\ell} < 80$ GeV & 216 & 3794 & 134 & 160 & 4088 & 3.3 \\
$110<m_{\text{recoil}}<150$ GeV & 215 & 2074 & 104 & 158 & 2336 & 4.3 \\
$|\eta_{\text{miss}}| < 2$ & 207 & 211 & 96 & 144 & 451 & 8.1 \\
MVA $> 0.6$ & 168 & 2 & 2 & 1 & 5 & 12.8 \\

\hline
Cumulative Efficiency (\%) & 70 & 0 & 0 & 0 & 0 & --- \\
\bottomrule
\end{tabular}}
\end{subtable}
&
\begin{subtable}{0.48\textwidth}
\centering
\caption{\chan{ee}{mumu}}
\resizebox{\textwidth}{!}{
\begin{tabular}{l r r r r r r}
\toprule
Selection & $S$ & $B_{\mathrm{ZZ}}$ & $B_{\mathrm{WW}}$ & $B_{ZH}$ & $B_{\rm tot}$ & $Z$ \\
\hline\hline
Preselection & 250 & 15139 & 1947 & 356 & 17442 & 1.9 \\
$80<m_Z<100$ GeV & 208 & 6325 & 68 & 149 & 6542 & 2.5 \\
$p_{\ell} < 80$ GeV & 208 & 3536 & 48 & 149 & 3733 & 3.3 \\
$110<m_{\text{recoil}}<150$ GeV & 207 & 2000 & 38 & 146 & 2184 & 4.2 \\
$|\eta_{\text{miss}}| < 2$ & 200 & 195 & 36 & 133 & 363 & 8.4 \\
MVA $> 0.6$ & 165 & 0 & 2 & 1 & 4 & 12.7 \\

\hline
Cumulative Efficiency (\%) & 66 & 0 & 0 & 0 & 0 & --- \\
\bottomrule
\end{tabular}}
\end{subtable}
\\[2ex]

\begin{subtable}{0.48\textwidth}
\centering
\caption{\chan{mumu}{mumu}}
\resizebox{\textwidth}{!}{
\begin{tabular}{l r r r r r r}
\toprule
Selection & $S$ & $B_{\mathrm{ZZ}}$ & $B_{\mathrm{WW}}$ & $B_{ZH}$ & $B_{\rm tot}$ & $Z$ \\
\hline\hline
Preselection & 235 & 7358 & 407 & 162 & 7926 & 2.6 \\
$80<m_Z<100$ GeV & 217 & 5635 & 91 & 138 & 5864 & 2.8 \\
$p_{\ell} < 80$ GeV & 216 & 2989 & 53 & 137 & 3179 & 3.7 \\
$110<m_{\text{recoil}}<150$ GeV & 213 & 1551 & 45 & 135 & 1731 & 4.8 \\
$|\eta_{\text{miss}}| < 2$ & 206 & 267 & 43 & 124 & 435 & 8.1 \\
MVA $> 0.6$ & 161 & 4 & 3 & 1 & 8 & 12.4 \\

\hline
Cumulative Efficiency (\%) & 69 & 0 & 1 & 1 & 0 & --- \\
\bottomrule
\end{tabular}}
\end{subtable}
&
\begin{subtable}{0.48\textwidth}
\centering
\caption{\chan{ee}{ee}}
\resizebox{\textwidth}{!}{
\begin{tabular}{l r r r r r r}
\toprule
Selection & $S$ & $B_{\mathrm{ZZ}}$ & $B_{\mathrm{WW}}$ & $B_{ZH}$ & $B_{\rm tot}$ & $Z$ \\
\hline\hline
Preselection & 253 & 7566 & 1904 & 177 & 9647 & 2.5 \\
$80<m_Z<100$ GeV & 218 & 5432 & 314 & 143 & 5889 & 2.8 \\
$p_{\ell} < 80$ GeV & 218 & 2997 & 198 & 142 & 3337 & 3.7 \\
$110<m_{\text{recoil}}<150$ GeV & 214 & 1610 & 164 & 140 & 1914 & 4.6 \\
$|\eta_{\text{miss}}| < 2$ & 206 & 291 & 156 & 129 & 576 & 7.4 \\
MVA $> 0.6$ & 160 & 5 & 3 & 2 & 10 & 12.3 \\

\hline
Cumulative Efficiency (\%) & 63 & 0 & 0 & 1 & 0 & --- \\
\bottomrule
\end{tabular}}
\end{subtable}
\\[2ex]

\begin{subtable}{0.48\textwidth}
\centering
\caption{\chan{mumu}{emu}}
\resizebox{\textwidth}{!}{
\begin{tabular}{l r r r r r r}
\toprule
Selection & $S$ & $B_{\mathrm{ZZ}}$ & $B_{\mathrm{WW}}$ & $B_{ZH}$ & $B_{\rm tot}$ & $Z$ \\
\hline\hline
Preselection & 478 & 6238 & 853 & 241 & 7332 & 5.4 \\
$80<m_Z<100$ GeV & 435 & 3537 & 41 & 189 & 3767 & 6.7 \\
$p_{\ell} < 80$ GeV & 434 & 1903 & 20 & 188 & 2111 & 8.6 \\
$110<m_{\text{recoil}}<150$ GeV & 430 & 531 & 19 & 186 & 737 & 12.6 \\
$|\eta_{\text{miss}}| < 2$ & 416 & 435 & 19 & 180 & 634 & 12.8 \\
MVA $> 0.6$ & 377 & 12 & 5 & 1 & 18 & 19.0 \\

\hline

Cumulative Efficiency (\%) & 79 & 0 & 0 & 0 & 0 & --- \\
\bottomrule
\end{tabular}}
\end{subtable}
&
\begin{subtable}{0.48\textwidth}
\centering
\caption{\chan{ee}{emu}}
\resizebox{\textwidth}{!}{
\begin{tabular}{l r r r r r r}
\toprule
Selection & $S$ & $B_{\mathrm{ZZ}}$ & $B_{\mathrm{WW}}$ & $B_{ZH}$ & $B_{\rm tot}$ & $Z$ \\
\hline\hline
Preselection & 498 & 6359 & 3346 & 254 & 9958 & 4.9 \\
$80<m_Z<100$ GeV & 419 & 3409 & 111 & 183 & 3703 & 6.5 \\
$p_{\ell} < 80$ GeV & 419 & 1858 & 56 & 183 & 2096 & 8.4 \\
$110<m_{\text{recoil}}<150$ GeV & 414 & 553 & 54 & 181 & 788 & 11.9 \\
$|\eta_{\text{miss}}| < 2$ & 401 & 461 & 53 & 175 & 689 & 12.1 \\
MVA $> 0.6$ & 359 & 13 & 3 & 1 & 17 & 18.5 \\
\hline

Cumulative Efficiency (\%) & 72 & 0 & 0 & 0 & 0 & --- \\
\bottomrule
\end{tabular}}
\end{subtable}

\end{tabular}
\end{table*}

\begin{table*}[htbp]
\centering
\caption{Event yields and significance for all channels at $\sqrt{s}=365$ GeV after applying selection conditions. Here, $B_{\rm{other}}$ includes processes such as $ZH$(Background) and $t\bar{t}$.}
\label{tab:cutFlow365}
\begin{tabular}{cc}

\begin{subtable}{0.48\textwidth}
\centering
\caption{\chan{mumu}{ee}}
\resizebox{\textwidth}{!}{
\begin{tabular}{l r r r r r r}
\toprule
Selection & $S$ & $B_{\mathrm{ZZ}}$ & $B_{\mathrm{WW}}$ & $B_{\rm{other}}$ & $B_{\rm tot}$ & $Z$ \\
\hline
\hline

Preselection & 47 & 2760 & 765 & 3460 & 6985 & 0.6 \\
$80<m_Z<100$ GeV & 41 & 1268 & 51 & 114 & 1433 & 1.1 \\
$110<m_{\text{recoil}}<150$ GeV & 39 & 268 & 47 & 105 & 420 & 1.8 \\
$|\eta_{\text{miss}}| < 2$ & 30 & 43 & 8 & 18 & 68 & 3.0 \\
MVA $> 0.6$ & 23 & 1 & 0 & 0 & 1 & 4.7 \\

\hline
Cumulative Efficiency (\%) & 49 & 0 & 0 & 0 & 0 & --- \\
\bottomrule
\end{tabular}}
\end{subtable}
&
\begin{subtable}{0.48\textwidth}
\centering
\caption{\chan{ee}{mumu}}
\resizebox{\textwidth}{!}{
\begin{tabular}{l r r r r r r}
\toprule
Selection & $S$ & $B_{\mathrm{ZZ}}$ & $B_{\mathrm{WW}}$ & $B_{\rm{other}}$ & $B_{\rm tot}$ & $Z$ \\
\hline\hline
Preselection & 75 & 2760 & 765 & 3460 & 6985 & 0.9 \\
$80<m_Z<100$ GeV & 39 & 1151 & 14 & 110 & 1274 & 1.1 \\
$110<m_{\text{recoil}}<150$ GeV & 38 & 230 & 13 & 103 & 346 & 1.9 \\
$|\eta_{\text{miss}}| < 2$ & 28 & 41 & 2 & 15 & 58 & 3.0 \\
MVA $> 0.6$ & 20 & 0 & 0 & 0 & 1 & 4.4 \\

\hline
Cumulative Efficiency (\%) & 27 & 0 & 0 & 0 & 0 & --- \\
\bottomrule
\end{tabular}}
\end{subtable}
\\[2ex]

\begin{subtable}{0.48\textwidth}
\centering
\caption{\chan{mumu}{mumu}}
\resizebox{\textwidth}{!}{
\begin{tabular}{l r r r r r r}
\toprule
Selection & $S$ & $B_{\mathrm{ZZ}}$ & $B_{\mathrm{WW}}$ & $B_{\rm{other}}$ & $B_{\rm tot}$ & $Z$ \\
\hline\hline
Preselection & 45 & 1289 & 162 & 884 & 2336 & 0.9 \\
$80<m_Z<100$ GeV & 41 & 1008 & 38 & 107 & 1153 & 1.2 \\
$110<m_{\text{recoil}}<150$ GeV & 39 & 271 & 36 & 100 & 408 & 1.9 \\
$|\eta_{\text{miss}}| < 2$ & 28 & 41 & 4 & 15 & 60 & 3.0 \\
MVA $> 0.6$ & 21 & 1 & 1 & 0 & 2 & 4.4 \\

\hline
Cumulative Efficiency (\%) & 47 & 0 & 1 & 0 & 0 & --- \\
\bottomrule
\end{tabular}}
\end{subtable}
&
\begin{subtable}{0.48\textwidth}
\centering
\caption{\chan{ee}{ee}}
\resizebox{\textwidth}{!}{
\begin{tabular}{l r r r r r r}
\toprule
Selection & $S$ & $B_{\mathrm{ZZ}}$ & $B_{\mathrm{WW}}$ & $B_{\rm{other}}$ & $B_{\rm tot}$ & $Z$ \\
\hline\hline
Preselection & 78 & 1394 & 787 & 1041 & 3222 & 1.3 \\
$80<m_Z<100$ GeV & 50 & 1009 & 150 & 129 & 1288 & 1.4 \\
$110<m_{\text{recoil}}<150$ GeV & 48 & 282 & 142 & 119 & 543 & 2.0 \\
$|\eta_{\text{miss}}| < 2$ & 28 & 50 & 22 & 15 & 87 & 2.6 \\
MVA $> 0.6$ & 17 & 0 & 0 & 0 & 1 & 4.1 \\

\hline
Cumulative Efficiency (\%) & 22 & 0 & 0 & 0 & 0 & --- \\
\bottomrule
\end{tabular}}
\end{subtable}
\\[2ex]

\begin{subtable}{0.48\textwidth}
\centering
\caption{\chan{mumu}{emu}}
\resizebox{\textwidth}{!}{
\begin{tabular}{l r r r r r r}
\toprule
Selection & $S$ & $B_{\mathrm{ZZ}}$ & $B_{\mathrm{WW}}$ & $B_{\rm{other}}$ & $B_{\rm tot}$ & $Z$ \\
\hline\hline
Preselection & 92 & 1095 & 296 & 3235 & 4626 & 1.3 \\
$80<m_Z<100$ GeV & 82 & 627 & 27 & 205 & 859 & 2.7 \\
$110<m_{\text{recoil}}<150$ GeV & 79 & 503 & 25 & 197 & 724 & 2.8 \\
$|\eta_{\text{miss}}| < 2$ & 58 & 76 & 4 & 25 & 105 & 4.6 \\
MVA $> 0.6$ & 49 & 3 & 1 & 0 & 4 & 6.7 \\

\hline

Cumulative Efficiency (\%) & 53 & 0 & 0 & 0 & 0 & --- \\
\bottomrule
\end{tabular}}
\end{subtable}
&
\begin{subtable}{0.48\textwidth}
\centering
\caption{\chan{ee}{emu}}
\resizebox{\textwidth}{!}{
\begin{tabular}{l r r r r r r}
\toprule
Selection & $S$ & $B_{\mathrm{ZZ}}$ & $B_{\mathrm{WW}}$ & $B_{\rm{other}}$ & $B_{\rm tot}$ & $Z$ \\
\hline\hline
Preselection & 150 & 1149 & 1327 & 3516 & 5991 & 1.9 \\
$80<m_Z<100$ GeV & 89 & 600 & 82 & 207 & 889 & 2.8 \\
$110<m_{\text{recoil}}<150$ GeV & 85 & 485 & 75 & 197 & 756 & 2.9 \\
$|\eta_{\text{miss}}| < 2$ & 55 & 86 & 12 & 24 & 123 & 4.1 \\
MVA $> 0.6$ & 43 & 2 & 0 & 0 & 2 & 6.4 \\
\hline
Cumulative Efficiency (\%) & 29 & 0 & 0 & 0 & 0 & --- \\
\bottomrule
\end{tabular}}
\end{subtable}

\end{tabular}
\end{table*}

We begin the selection criteria by requiring at least one pair of leptons with invariant mass consistent with the mass of the $Z$ boson. In all channels, this selection criteria reduces the background from the $WW$ process, as expected, due to the absence of a $Z$ boson. Moreover, an upper selection criteria on the lepton momenta is useful to remove background events from the $ZZ$ and $WW$ processes. 

\begin{figure*}[tbp]
        \subfloat[\centering \chan{mumu}{ee}]{\includegraphics[width=0.32\linewidth]{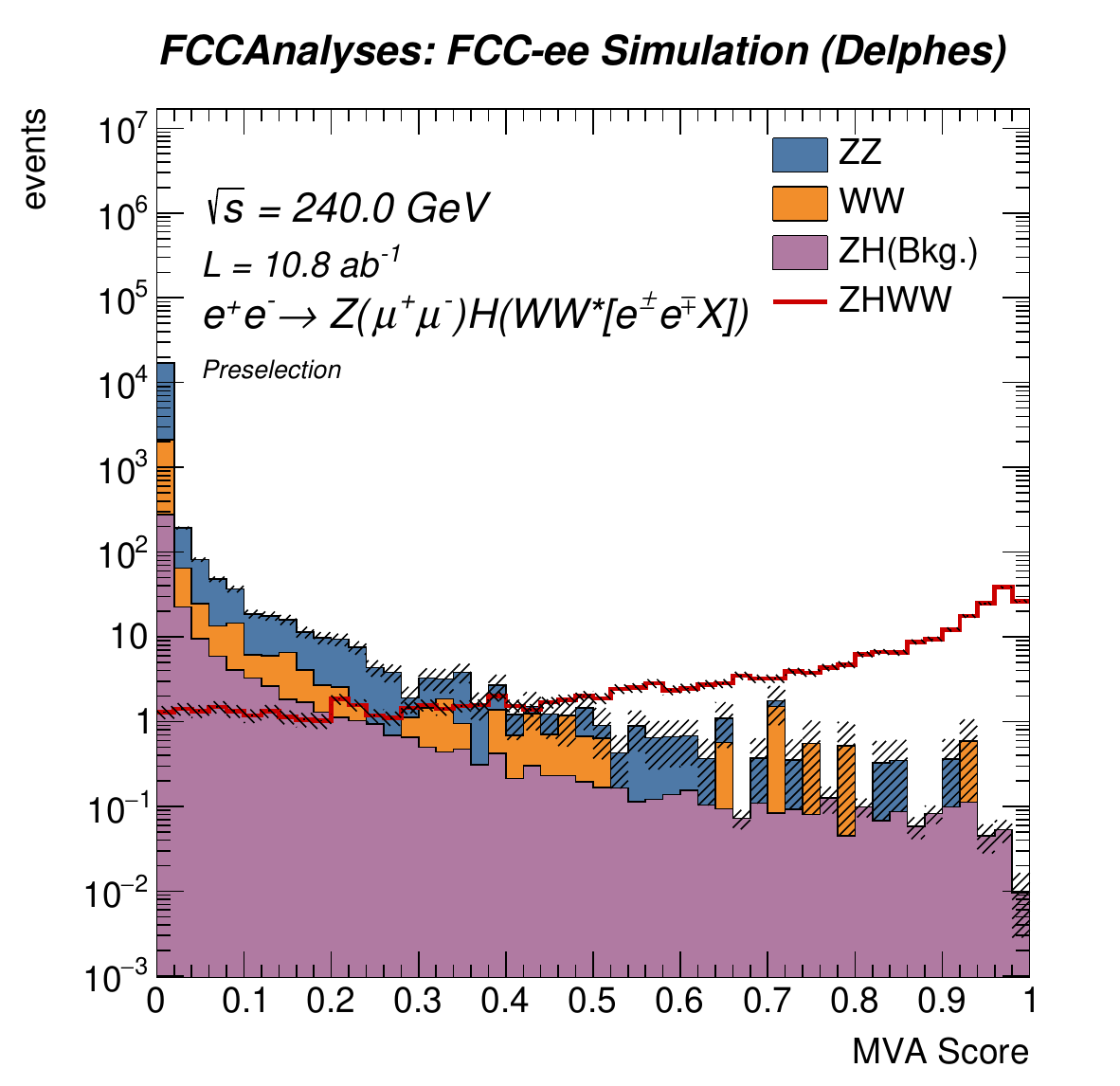}}
        \subfloat[\centering \chan{ee}{mumu}]{\includegraphics[width=0.32\linewidth]{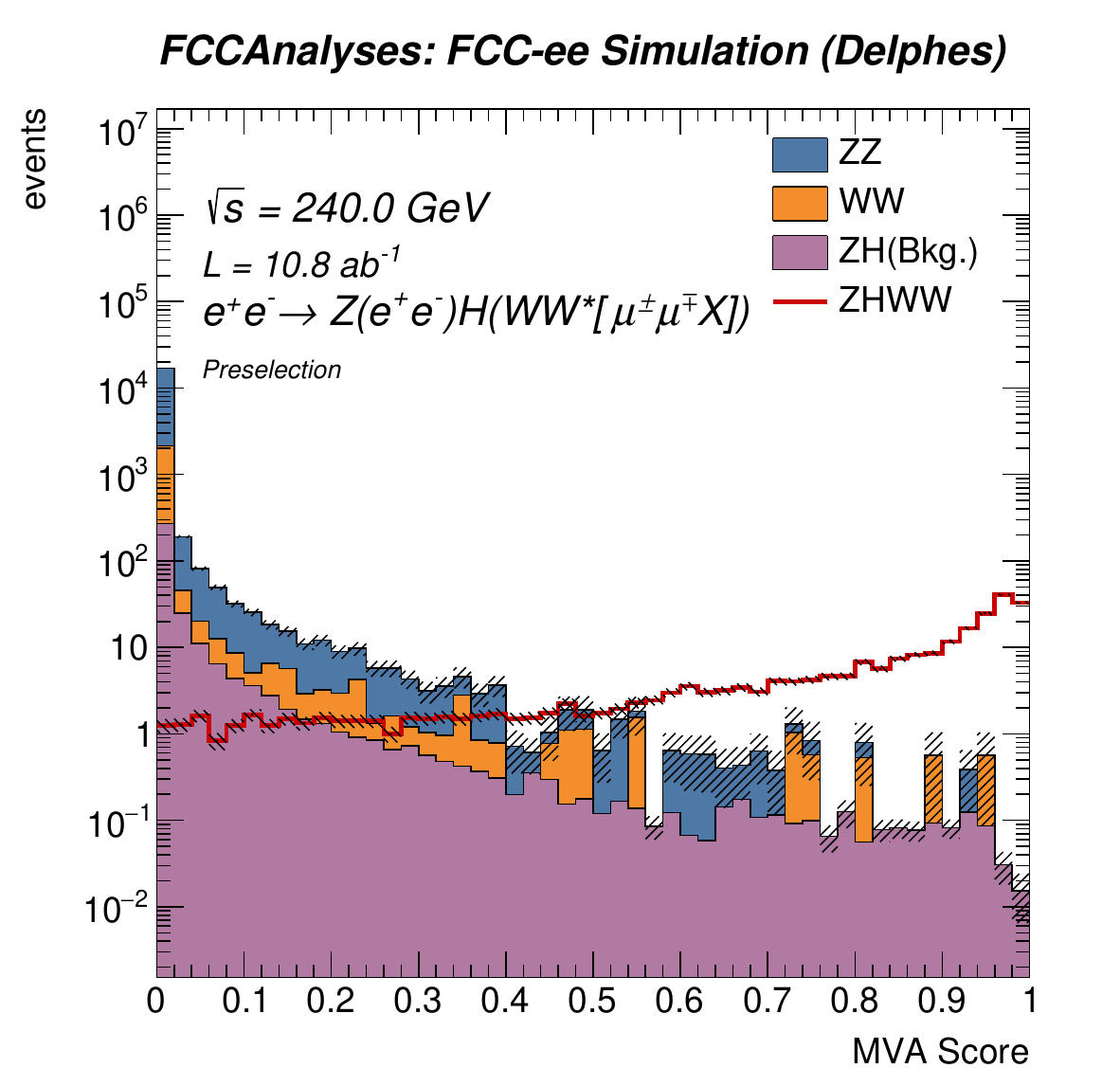}}
        \subfloat[\centering \chan{mumu}{mumu}]{\includegraphics[width=0.32\linewidth]{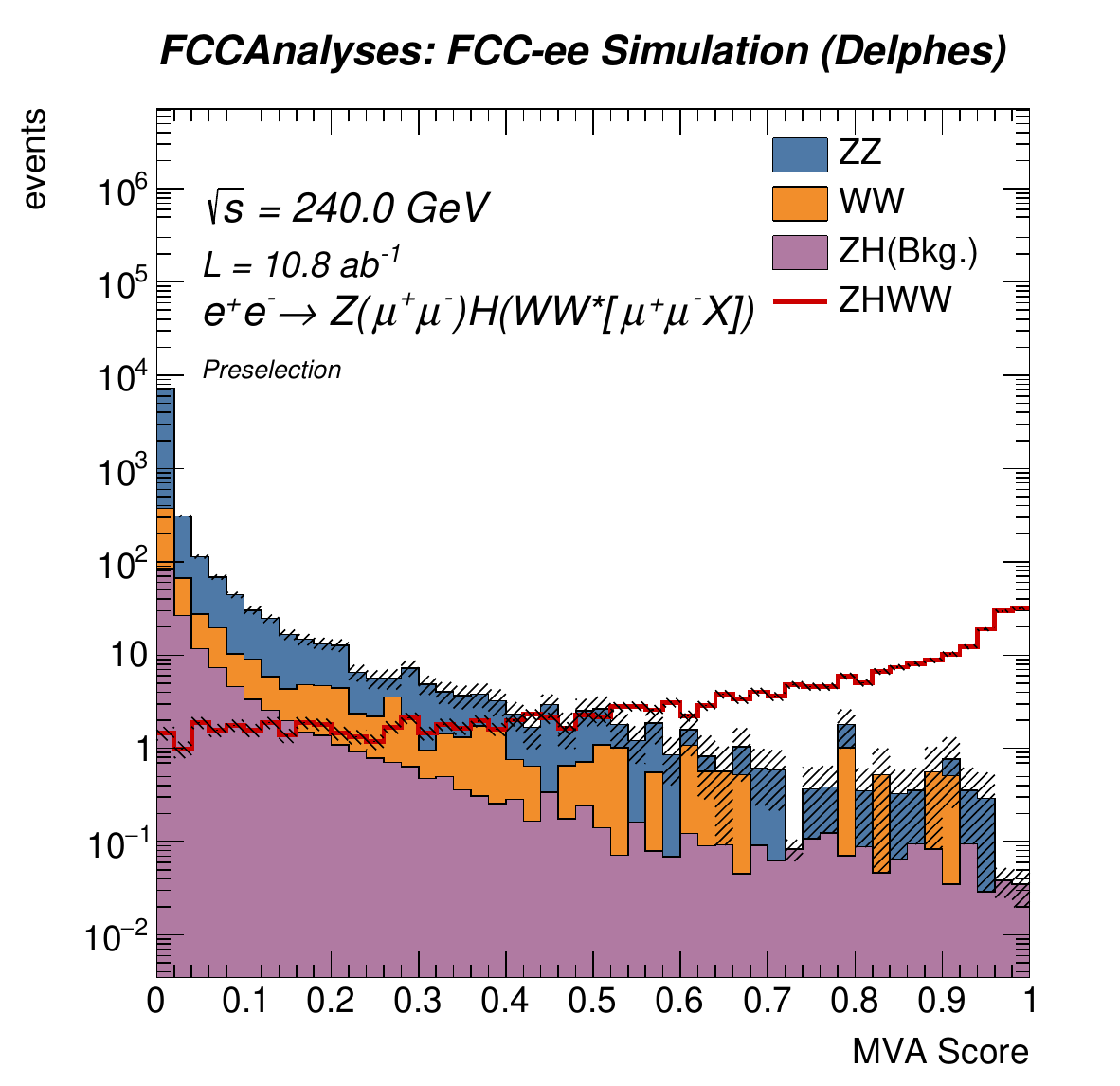}}
        \qquad
        \subfloat[\centering \chan{ee}{ee}]{\includegraphics[width=0.32\linewidth]{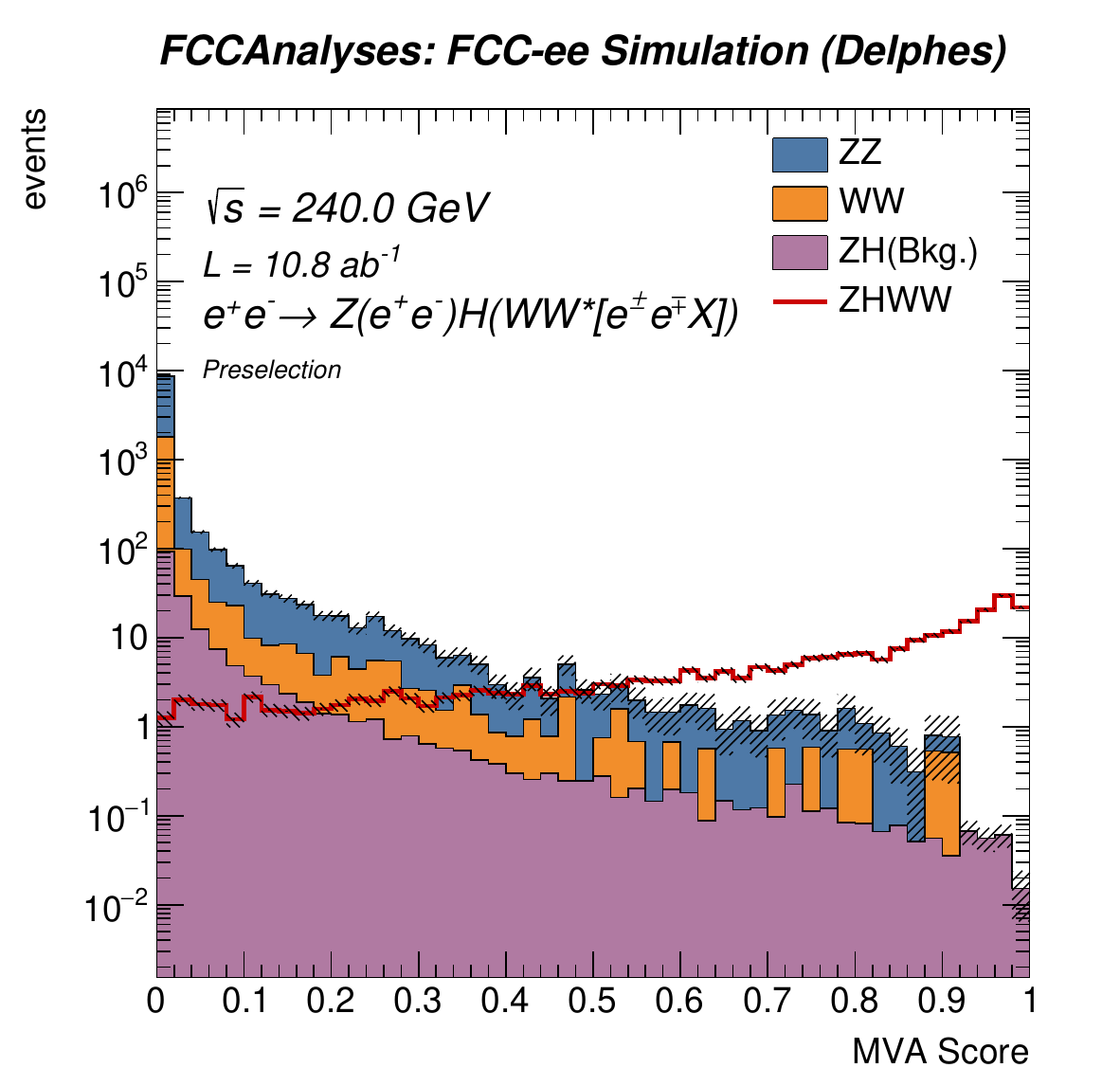}}
        \subfloat[\centering \chan{mumu}{emu}]{\includegraphics[width=0.32\linewidth]{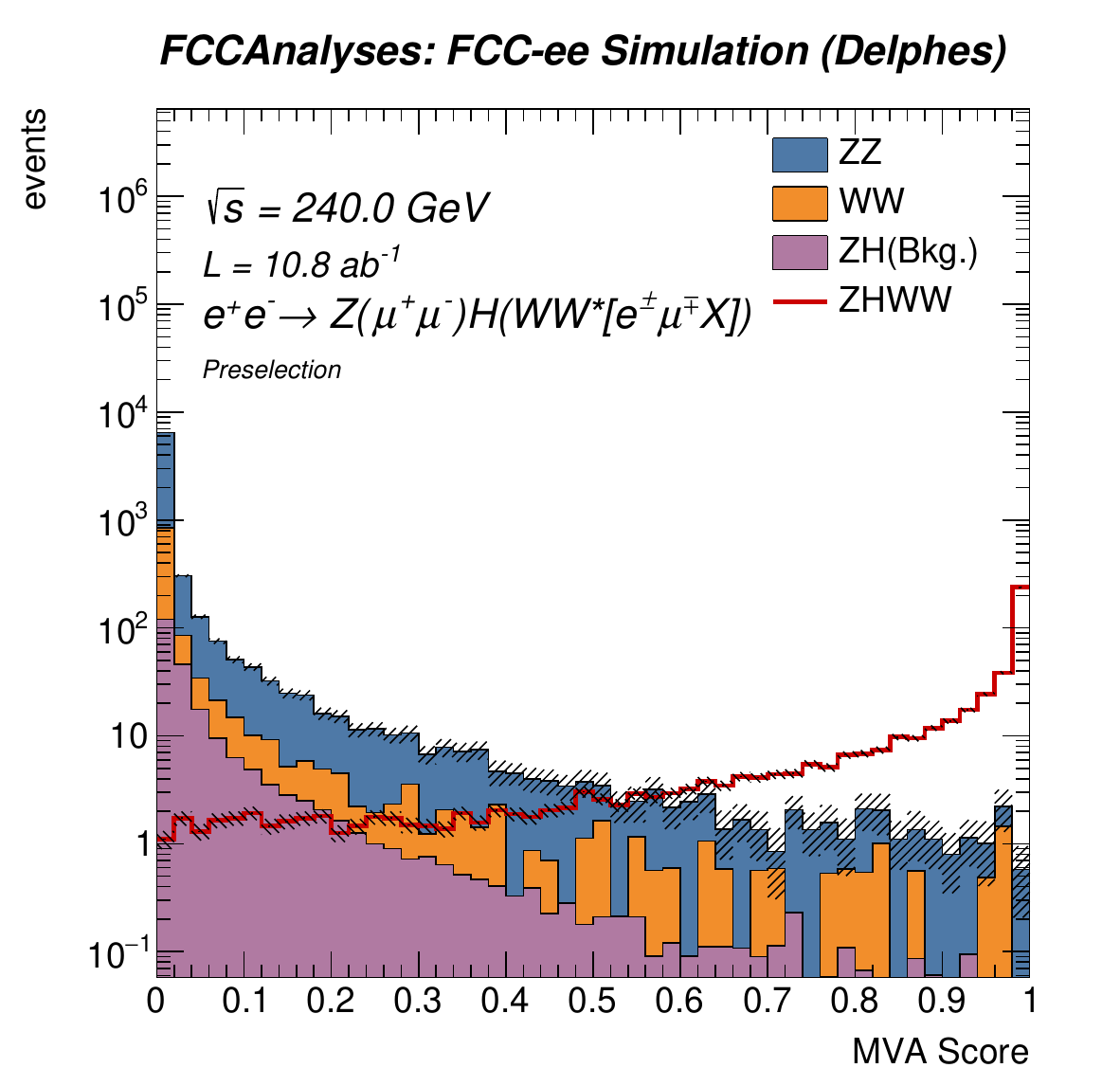}}
        \subfloat[\centering \chan{ee}{emu}]{\includegraphics[width=0.32\linewidth]{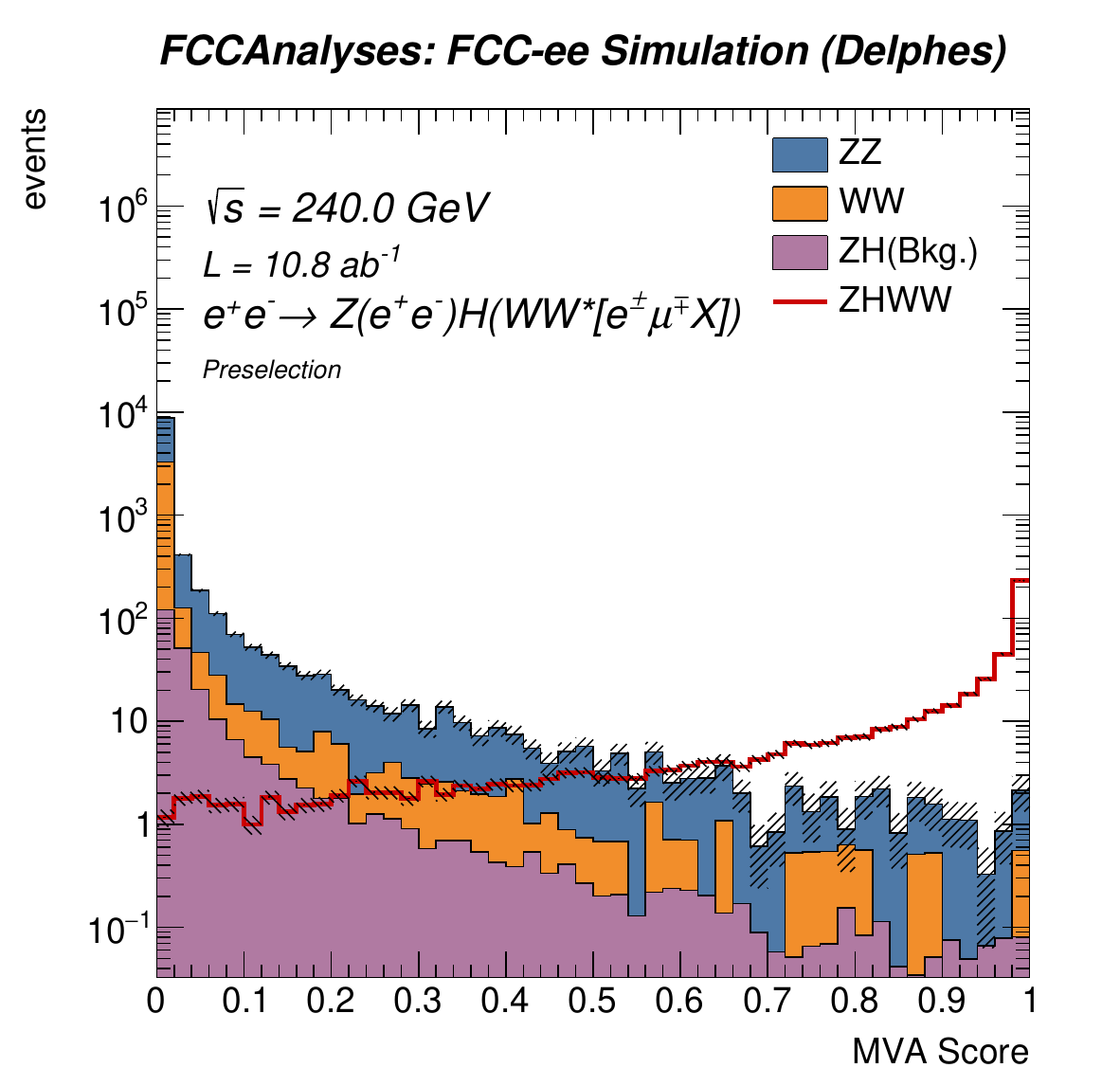}}
        \caption{MVA score for the \chan{mumu}{ee}, \chan{ee}{mumu}, \chan{mumu}{mumu}, \chan{ee}{ee}, \chan{mumu}{emu},  \chan{ee}{emu} channels for $\sqrt{s}=240$ GeV. The uncertainty associated with the MC simulation is shown by the shaded area.}
        \label{fig:mva_score240}
\end{figure*}

\begin{figure*}[tbp]
        \subfloat[\centering \chan{mumu}{ee}]{\includegraphics[width=0.32\linewidth]{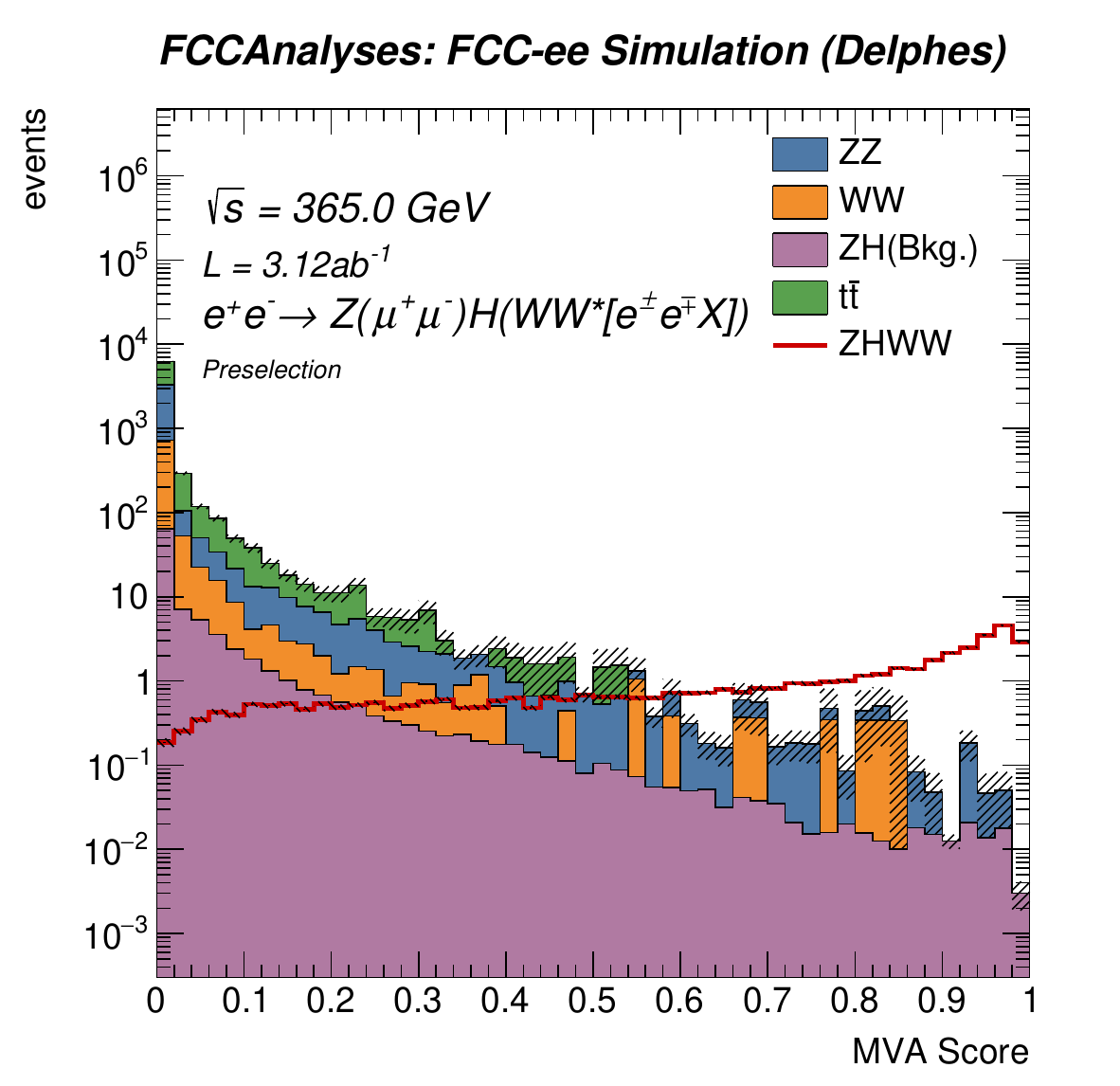}}
        \subfloat[\centering \chan{ee}{mumu}]{\includegraphics[width=0.32\linewidth]{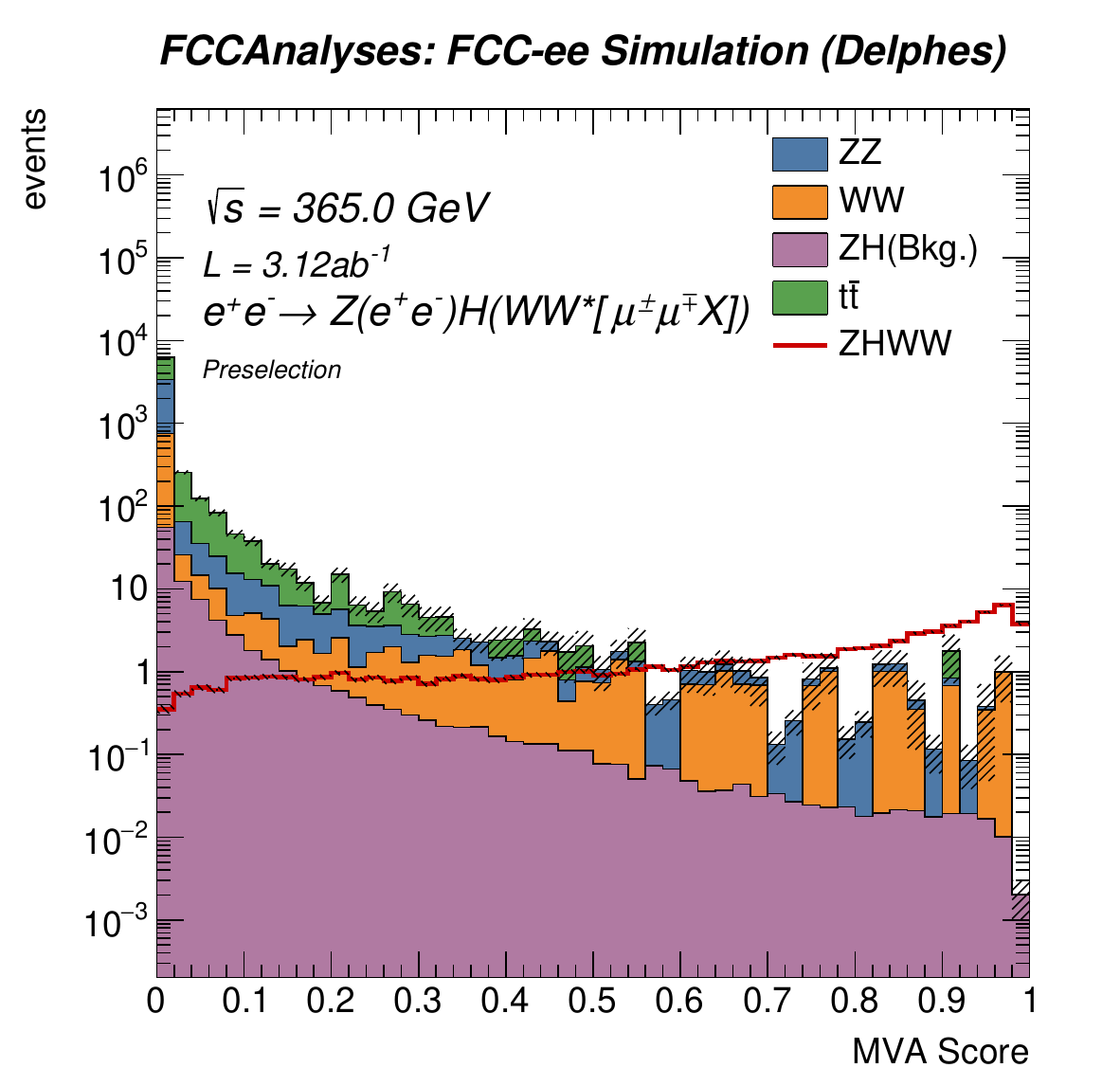}}
        \subfloat[\centering \chan{mumu}{mumu}]{\includegraphics[width=0.32\linewidth]{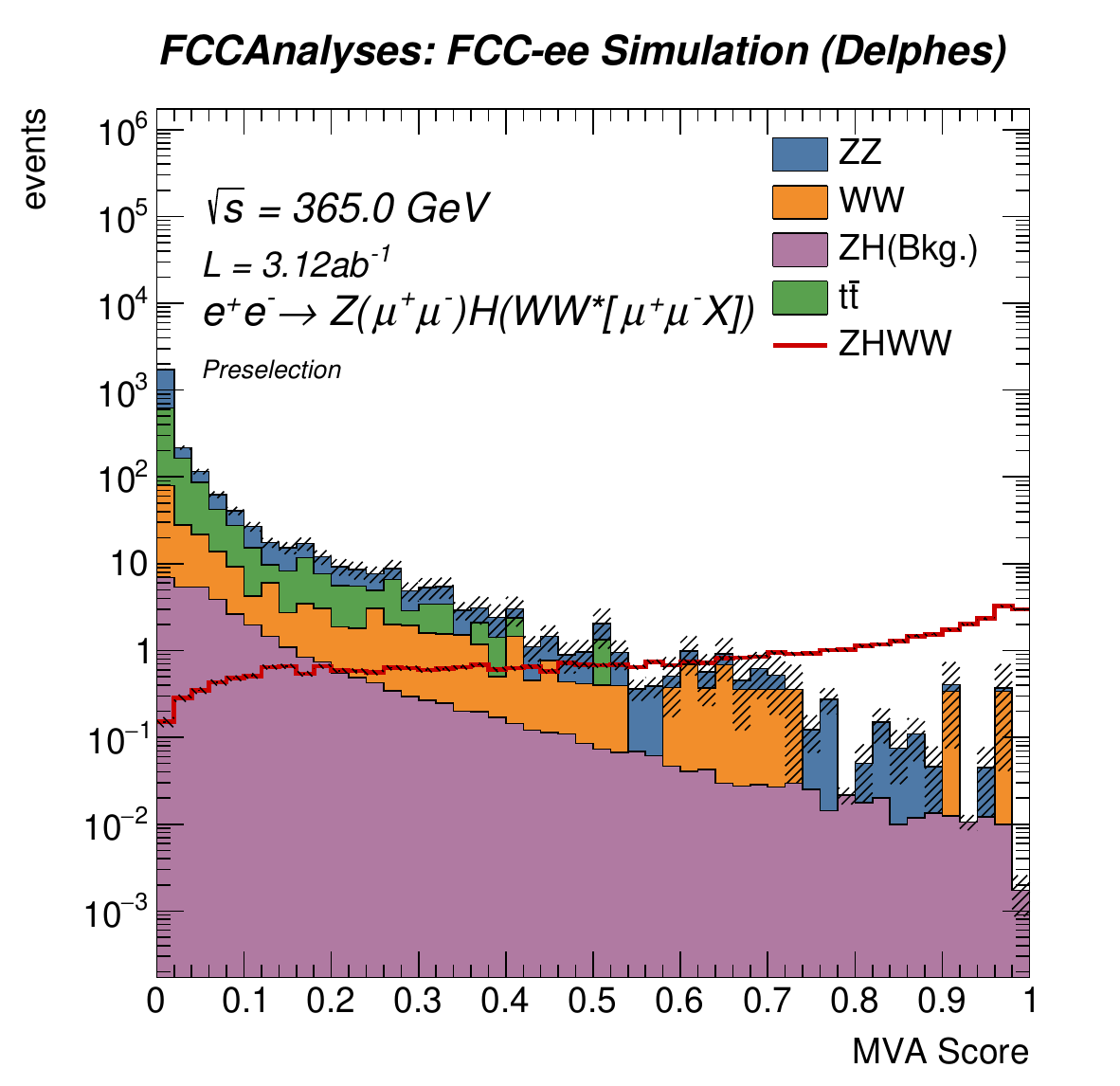}}
        \qquad
        \subfloat[\centering \chan{ee}{ee}]{\includegraphics[width=0.32\linewidth]{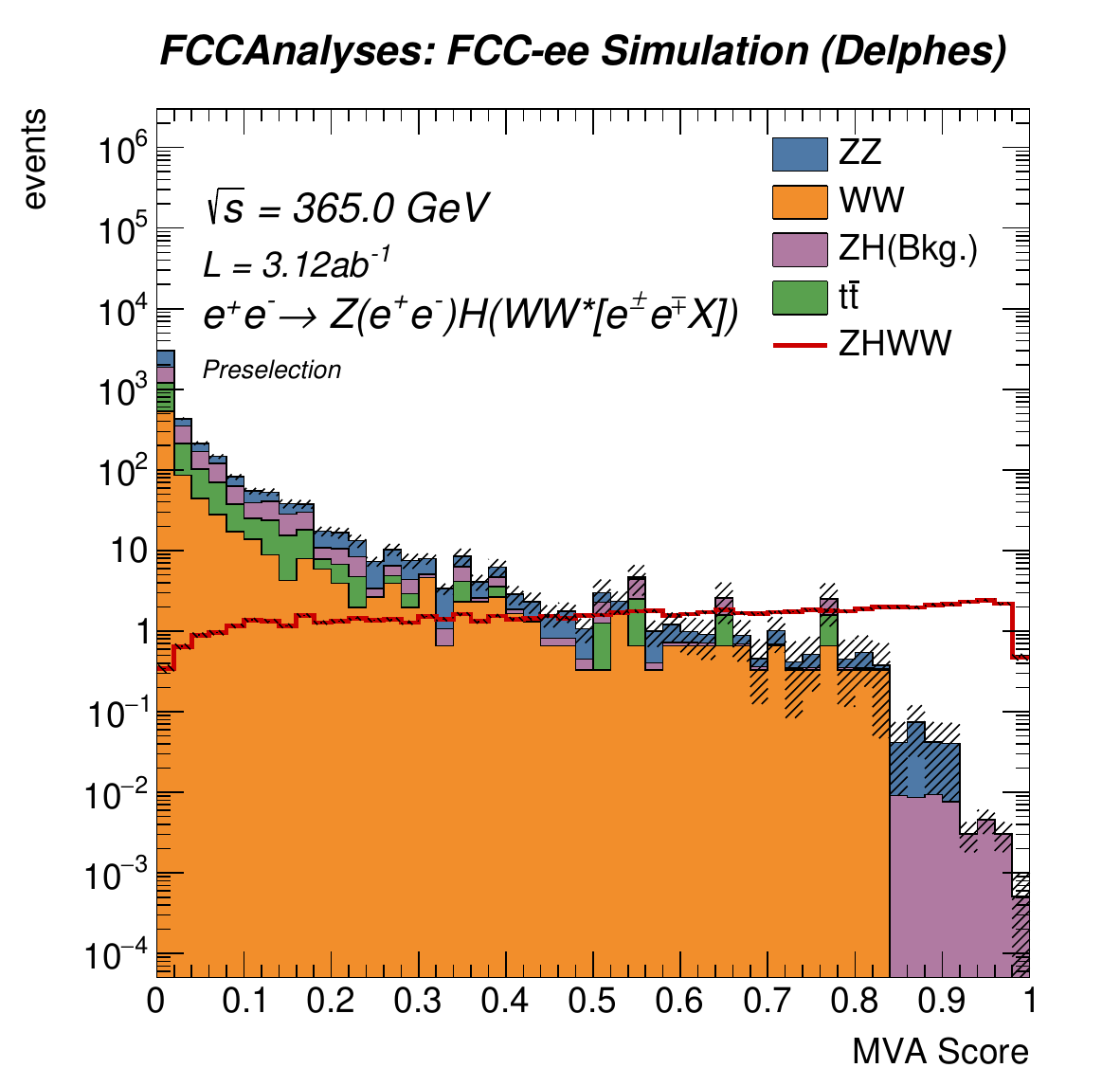}}
        \subfloat[\centering \chan{mumu}{emu}]{\includegraphics[width=0.32\linewidth]{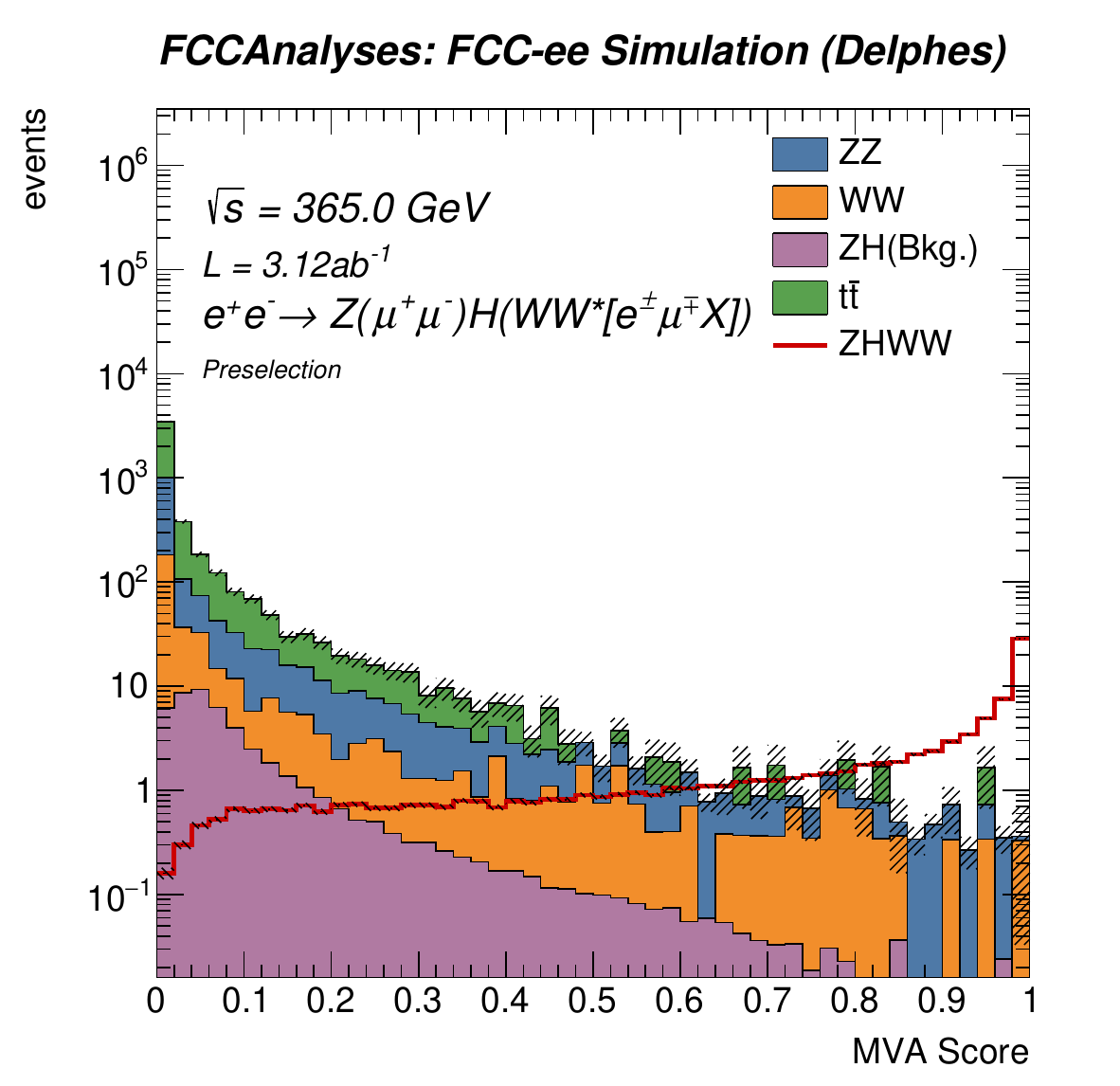}}
        \subfloat[\centering \chan{ee}{emu}]{\includegraphics[width=0.32\linewidth]{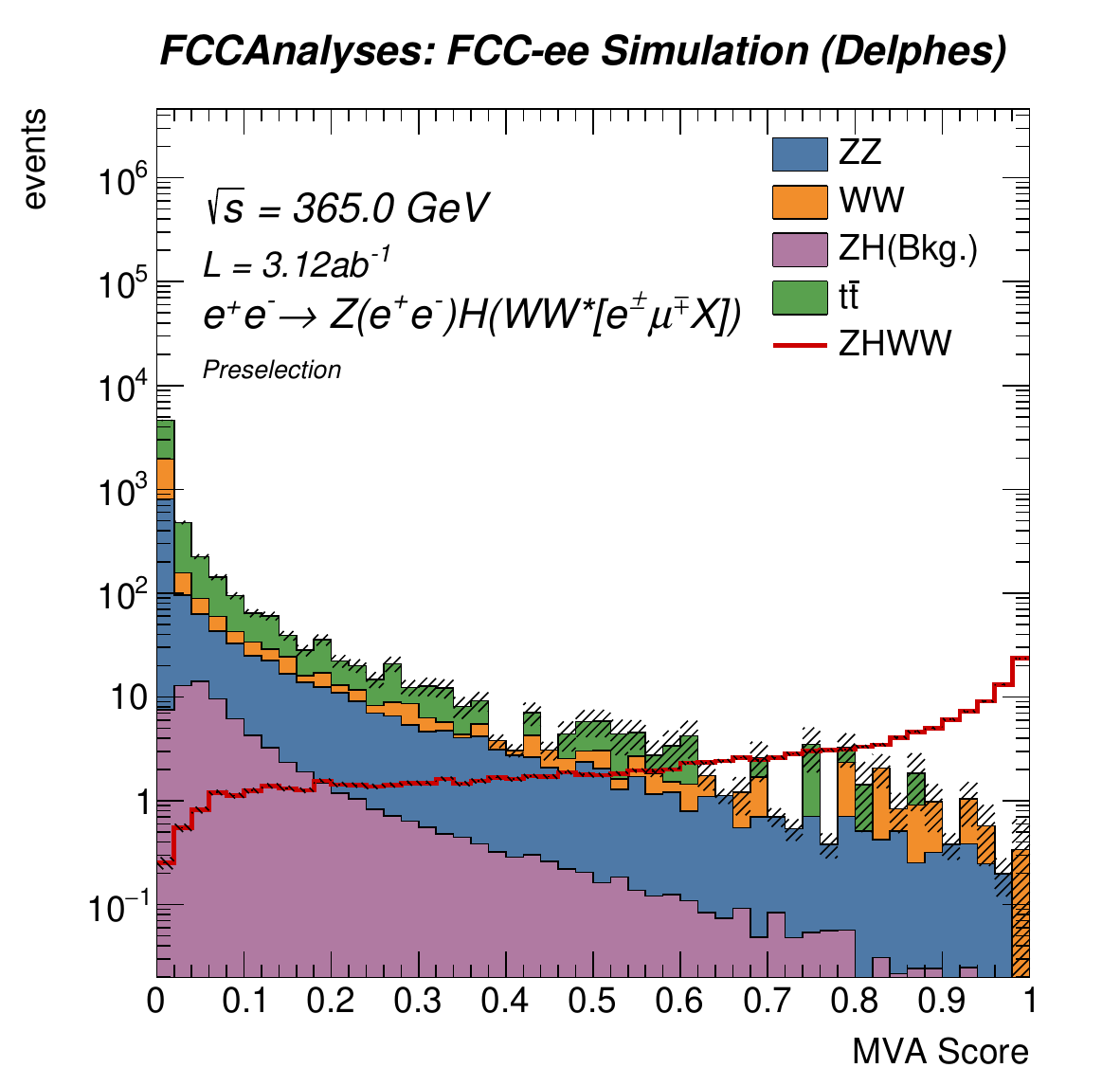}}
        \caption{MVA score for the \chan{mumu}{ee}, \chan{ee}{mumu}, \chan{mumu}{mumu}, \chan{ee}{ee}, \chan{mumu}{emu},  \chan{ee}{emu} channels for $\sqrt{s}=365$ GeV. The uncertainty associated with the MC simulation is shown by the shaded area.}
        \label{fig:mva_score365}
\end{figure*}

For the signal sample, the recoil mass is expected to be close to the Higgs boson mass. This additional observable can be used to reject events not consistent with the presence of the Higgs boson as an intermediate quantum state. Furthermore, we select events satisfying $|\eta_{\rm miss}| < 2.0$ motivated by the presence of neutrinos in final state in this region. 

The results from the selection based analysis show that we achieve $8.1\sigma$ and $8.4\sigma$ in \chan{mumu}{ee} and \chan{ee}{mumu} channels, respectively. For these channels, the background is mainly dominated by the $e^+ e^- \rightarrow ZH(\rightarrow ZZ)$ process where two $Z$ bosons decay into leptons and one of the $Z$ bosons decays to a pair of neutral leptons. In the four muons and four electrons channels, \chan{mumu}{mumu} and \chan{ee}{ee} give a significance of $8.2\sigma$ and $7.4\sigma$, respectively. Here, the background is dominated by $e^+ e^- \rightarrow ZZ$ which yields exactly the same visible final state, followed by contributions from $ZH$ processes. Across all channels, the highest significance, of $12.1\sigma$ and $12.8\sigma$ is noted in the \chan{ee}{emu} and \chan{mumu}{emu} channels, respectively.

We also developed a Machine Learning based approach to improve sensitivity when discriminating the signal from the background. We used a Boosted Decision Tree (BDT) from XGBoost package~\cite{Chen:2016:XGBoost} for a signal versus background classification task. A set of both, low-level features, including the four-momenta of the leptons, missing momenta and the associated pseudorapidity, visible energy and transverse visible momenta, and high level features such as the invariant mass of the dilepton pairs, angular distance between leptons, mass and energy of Z boson and the recoil mass are used. We also evaluated the fraction of missing energy over the total energy.  Angular variables based on the $\Delta R$ distance between pairs of all possible leptons is evaluated using the following equation:

\begin{equation}
    \Delta R(\ell_1, \ell_2) = \sqrt{\Delta\phi^2(\ell_1, \ell_2) + \Delta\eta^2(\ell_1, \ell_2) } 
\end{equation}

We used 44 features to train the BDT. The parameters defining the BDT model are summarised in \autoref{tab:hyperparam}. The set of parameters are chosen so as not to overfit or underfit the training dataset. The binary logistic function is used for binary classification.  The maximum depth of each tree is set to five to prevent overfitting. The number of estimators is set to 1000, while early stopping rounds are set to 50. Several parameters, such as regularisation, were employed to further reduce overfitting. The preselected events are used to train and test the model. These events are split into two sets: training and testing phase, with a ratio of 3:1. We train the model for each given channel separately, and the results obtained are saved using the \textsc{Root Tmva} feature~\cite{Brun:1997pa} in order to assess the impact on the entire set of event samples. The outcome of the MVA on the preselected events is presented in \autoref{fig:mva_score240} for all six channels at 240 GeV.

\begin{table*}[tbp]

\centering
\caption{XGBoost hyperparameters used for training the Machine Learning Model for $\sqrt{s}=240, 365$ GeV.}
\label{tab:hyperparam}
\begin{tabular}{l l l}
\toprule
\textbf{Parameter} & \textbf{Value} & \textbf{Description} \\
\hline\hline
Objective & binary:logistic & Binary classification \\
Evaluation Metric & logloss & Loss function for evaluation \\
Max Depth & 5 & Maximum depth of tree \\
Learning Rate & 0.03 & Learning Rate for Model \\
Number of Estimators & 1000 & Number of Trees \\
Subsample & 0.6 & Fraction of data used per tree \\
Column Sample by Tree & 0.6 & Fraction of features used per tree \\
Gamma & 1.0 & Minimum loss reduction to split \\
Min Child Weight & 5 & Minimum sum of instance weight in child \\
L1 Regularization ($\alpha$) & 0.5 & L1 regularization term \\
L2 Regularization ($\lambda$) & 3.0 & L2 regularization term \\
Tree Method & hist & Histogram-based tree  \\
Early Stopping Rounds & 50 & Stops the code if no improvement after 50 steps \\
\bottomrule
\end{tabular}

\end{table*}

Across all channels, we selected events satisfying MVA score greater than 0.6. This leads to an improvement in significance in all channels. The \chan{mumu}{ee} and \chan{ee}{mumu} channel sees an increase in significance by 58\% and 51\%, respectively. In the \chan{mumu}{mumu} a rise in significance by 53\% is noted while the \chan{ee}{ee} channels has an improvement of about 66\% in signal versus background discrimination. In the \chan{mumu}{emu} and \chan{ee}{emu} channels, the significance shows an improvement by about 48\% and 53\%, respectively.  This approach show an increase in significance to up to $\sim 12.3\sigma$ to $19\sigma$ in \chan{ee}{ee} and \chan{mumu}{emu} channels, respectively.

A similar approach is followed for analysing $\sqrt{s} = 365 ~\rm{GeV}$. In \autoref{tab:cutFlow365}, a set of selection criterion along with the expected yields are presented. The preselection criteria introduced in the previous section lead to the following significance: $0.6\sigma$ in \chan{mumu}{ee}; $0.9\sigma$ in \chan{ee}{mumu} and \chan{mumu}{mumu}; $1.3\sigma$ in \chan{ee}{ee} and \chan{mumu}{emu};  and $1.9\sigma$ in \chan{ee}{emu}. The selection criteria used for $\sqrt{s} = 240 ~\rm{GeV}$  analysis are applied with the exception of the selection on lepton momentum. We also trained BDT from \textsc{XGBoost} package, keeping the same network parameters as outlined in \autoref{tab:hyperparam}.

Applying the selection on MVA score greater than 0.6, leads to significance of $4.7\sigma$ in \chan{mumu}{ee}; $4.4\sigma$ in \chan{ee}{mumu} and \chan{mumu}{mumu}; $4.1\sigma$ in \chan{ee}{ee}. We note as before that the channels \chan{mumu}{emu} and \chan{ee}{emu} show the highest significance across channels corresponding to $6.7\sigma$ and $6.4\sigma$, respectively.


\clearpage

\section{Determining the Signal Strength}
\label{sec:sigstrength}

The signal strength for $\sigma_{ZH} \times Br(H\rightarrow WW)$ is obtained using the \textsc{CMS Combine} package~\cite{CMS:2024onh}. To measure this, we used the MVA score as the observable after applying all selection criteria, except the final selection on the MVA score. We obtain the pseudo-data by adding the signal and background events obtained after appropriate scaling with luminosity and cross-section. The background processes are assigned a 10\% normalisation uncertainty, while luminosity is assigned a 1\% uncertainty. The relative uncertainty for each channel is then evaluated. As each of the six channels are orthogonal to one another, we have also evaluated the combined relative uncertainty in the $Z(\ell\ell)H(WW\rightarrow \ell\ell \nu\nu)$. Our results are summarised in \autoref{tab:results240} for $\sqrt{s}=240 ~\rm{GeV}$ and in \autoref{tab:results365} for $\sqrt{s}=365 ~\rm{GeV}$. For $\sqrt{s}=240 ~\rm{GeV}$ the relative uncertainty is lowest in the \chan{mumu}{emu} channel at a value of 5.1\% followed by the \chan{ee}{emu} channel with an uncertainty at a value of 5.3\%. The highest relative uncertainty is found in the \chan{ee}{ee} channel, with a value of 7.2\%. 

For the $\sqrt{s}=240 ~\rm{GeV}$, the combined relative uncertainty obtained by combining the six channels within \textsc{CMS Combine} tool is found to be 2.9\% which, approximately, corresponds to a signal significance of $35.1\sigma$, assuming that the $\rm{uncertainty}~\propto~\rm{1/significance}$.

Owing to limited sample statistics for the $\sqrt{s} = 365~\rm{GeV}$, we find that the relative uncertainty is  $\sim 14\%$ for all channels. The combined uncertainty is found as 6.8\%. This corresponds to a signal significance of $14.7\sigma$.

\begin{table}[htbp]
    \centering
    \caption{Significance and relative uncertainty for the six channels considered in this analysis for $\sqrt{s} = 240 ~\rm{GeV}$.}
    \label{tab:results240}
    
    \resizebox{0.5\textwidth}{0.15\textwidth}{
    \begin{tabular}{lccc}
        \toprule
        \textbf{Channel} & \textbf{Relative uncertainty (\%)} & \textbf{Significance (\sig)} \\
        \hline\hline
        \chan{mumu}{ee}   & 6.9 & 12.8 \\
        \chan{ee}{mumu}   & 7.0 & 12.7 \\
        \midrule
        \chan{mumu}{mumu} & 7.1 & 12.4 \\
        \chan{ee}{ee}     & 7.2 & 11.8 \\
        \midrule
        \chan{mumu}{emu}  & 5.1 & 19.0 \\
        \chan{ee}{emu}    & 5.3 & 18.5 \\
        \midrule
        \textbf{Combined} & \textbf{2.9} & \textbf{$\sim$ 35.1} \\
        \bottomrule
    \end{tabular}
    }
\end{table}

\begin{table}[htbp]
    \centering
    \caption{Significance and relative uncertainty for the six channels considered in this analysis for $\sqrt{s} = 365 ~\rm{GeV}$.}
    \label{tab:results365}
    
    \resizebox{0.5\textwidth}{0.15\textwidth}{
    \begin{tabular}{lccc}
        \toprule
        \textbf{Channel} & \textbf{Relative uncertainty (\%)} & \textbf{Significance (\sig)} \\
        \hline\hline
        \chan{mumu}{ee}   & 14.5 & 4.7 \\
        \chan{ee}{mumu}   & 14.2 & 4.4 \\
        \midrule
        \chan{mumu}{mumu} & 14.2 & 4.4 \\
        \chan{ee}{ee}     & 14.1 & 4.1 \\
        \midrule
        \chan{mumu}{emu}  & 14.9 & 6.7 \\
        \chan{ee}{emu}    & 14.9 & 6.4 \\
        \midrule
        \textbf{Combined} & \textbf{6.8} & \textbf{$\sim$ 14.7} \\
        \bottomrule
    \end{tabular}
    }
\end{table}

\section{Conclusion}\label{sec:conclusion}

We have studied the prospects of measuring $Z(\ell\ell)H\rightarrow WW (\ell\ell\nu\nu)$ in the four lepton final state at the FCC-ee. The analysis was performed at two centre-of-mass energies, $\sqrt{s} = 240~\rm{GeV}$ and $\sqrt{s} = 365~\rm{GeV}$, considering an integrated luminosity of 10.8 \ab~and 3.12 \ab, respectively.

The analysis was carried out in six orthogonal channels defined according to the number of electrons and muons in the final states. The use of Multivariate Analysis Techniques by incorporating Boosted Decision Tree enhanced the signal over background significantly.

We obtain a combined relative uncertainty of 2.9\% on measuring $\sigma(e^+ e^- \rightarrow ZH) \times Br(H\rightarrow W^+W^-)$ at $\sqrt{s} = 240~\rm{GeV}$. This corresponds to a combined significance of $35.1\sigma$. On the other hand, at $\sqrt{s} = 365~\rm{GeV}$, the relative uncertainties are worse by a factor of 2-3 times. The corresponding combined relative uncertainty is 6.8\%, corresponding to a significance of $14.7\sigma$. The results indicate more precise measurements of $\sigma(e^+ e^- \rightarrow ZH) \times Br(H\rightarrow W^+W^-)$ are possible at $\sqrt{s} = 240~\rm{GeV}$ energy compared to when FCC-ee is running at $\sqrt{s} = 365~\rm{GeV}$ energy.

\bibliographystyle{apsrev4-2}
\bibliography{refs}

@Article{HIGG-2012-27,
    author         = "{ATLAS Collaboration}",
    title          = "{Observation of a new particle in the search for the Standard Model Higgs boson with the ATLAS detector at the LHC}",
    journal        = "Phys. Lett. B",
    volume         = "716",
    year           = "2012",
    pages          = "1",
    doi            = "10.1016/j.physletb.2012.08.020",
    reportNumber   = "CERN-PH-EP-2012-218",
    eprint         = "1207.7214",
    archivePrefix  = "arXiv",
    primaryClass   = "hep-ex"
}

@article{ParticleDataGroup:2024cfk,
    author = "Navas, S. and others",
    collaboration = "Particle Data Group",
    title = "{Review of particle physics}",
    doi = "10.1103/PhysRevD.110.030001",
    journal = "Phys. Rev. D",
    volume = "110",
    number = "3",
    pages = "030001",
    year = "2024"
}

@misc{FCC:2025lpp,
    author = "Benedikt, M. and others",
    collaboration = "FCC",
    title = "{Future Circular Collider Feasibility Study Report: Volume 1, Physics, Experiments, Detectors}",
    eprint = "2505.00272",
    archivePrefix = "arXiv",
    primaryClass = "hep-ex",
    reportNumber = "CERN-FCC-PHYS-2025-0002",
    month = "4",
    year = "2025"
}

@Article{CMS-HIG-12-028,
    author         = "{CMS Collaboration}",
    title          = "{Observation of a new boson at a mass of 125 GeV with the CMS experiment at the LHC}",
    journal        = "Phys. Lett. B",
    volume         = "716",
    year           = "2012",
    pages          = "30",
    doi            = "10.1016/j.physletb.2012.08.021",
    reportNumber   = "CERN-PH-EP-2012-220",
    eprint         = "1207.7235",
    archivePrefix  = "arXiv",
    primaryClass   = "hep-ex"
}

@article{Englert:1964et,
    author = "Englert, F. and Brout, R.",
    editor = "Taylor, J. C.",
    title = "{Broken Symmetry and the Mass of Gauge Vector Mesons}",
    doi = "10.1103/PhysRevLett.13.321",
    journal = "Phys. Rev. Lett.",
    volume = "13",
    pages = "321--323",
    year = "1964"
}

@article{Higgs:1964ia,
    author = "Higgs, Peter W.",
    title = "{Broken symmetries, massless particles and gauge fields}",
    doi = "10.1016/0031-9163(64)91136-9",
    journal = "Phys. Lett.",
    volume = "12",
    pages = "132--133",
    year = "1964"
}

@misc{IDEAStudyGroup:2025gbt,
    author = "Abbrescia, M. and others",
    collaboration = "IDEA Study Group",
    title = "{The IDEA detector concept for FCC-ee}",
    eprint = "2502.21223",
    archivePrefix = "arXiv",
    primaryClass = "physics.ins-det",
    reportNumber = "FERMILAB-PUB-25-0189-PPD",
    month = "February",
    year = "2025"
}

@article{Kilian:2007gr,
    author = "Kilian, Wolfgang and Ohl, Thorsten and Reuter, Jurgen",
    title = "{WHIZARD: Simulating Multi-Particle Processes at LHC and ILC}",
    eprint = "0708.4233",
    archivePrefix = "arXiv",
    primaryClass = "hep-ph",
    reportNumber = "DESY-11-126, EDINBURGH-2010-36, FR-PHENO-2010-037, SI-HEP-2010-18",
    doi = "10.1140/epjc/s10052-011-1742-y",
    journal = "Eur. Phys. J. C",
    volume = "71",
    pages = "1742",
    year = "2011"
}

@article{Sjostrand:2014zea,
    author = "Sj{\"o}strand, Torbj{\"o}rn and Ask, Stefan and Christiansen, Jesper R. and Corke, Richard and Desai, Nishita and Ilten, Philip and Mrenna, Stephen and Prestel, Stefan and Rasmussen, Christine O. and Skands, Peter Z.",
    title = "{An introduction to PYTHIA 8.2}",
    eprint = "1410.3012",
    archivePrefix = "arXiv",
    primaryClass = "hep-ph",
    reportNumber = "LU-TP-14-36, MCNET-14-22, CERN-PH-TH-2014-190, FERMILAB-PUB-14-316-CD, DESY-14-178, SLAC-PUB-16122",
    doi = "10.1016/j.cpc.2015.01.024",
    journal = "Comput. Phys. Commun.",
    volume = "191",
    pages = "159--177",
    year = "2015"
}

@article{Sjostrand:2006za,
    author = "Sjostrand, Torbjorn and Mrenna, Stephen and Skands, Peter Z.",
    title = "{PYTHIA 6.4 Physics and Manual}",
    eprint = "hep-ph/0603175",
    archivePrefix = "arXiv",
    reportNumber = "FERMILAB-PUB-06-052-CD-T, LU-TP-06-13",
    doi = "10.1088/1126-6708/2006/05/026",
    journal = "JHEP",
    volume = "05",
    pages = "026",
    year = "2006"
}

@article{deFavereau:2013fsa,
    author = "de Favereau, J. and Delaere, C. and Demin, P. and Giammanco, A. and Lema{\^\i}tre, V. and Mertens, A. and Selvaggi, M.",
    collaboration = "DELPHES 3",
    title = "{DELPHES 3, A modular framework for fast simulation of a generic collider experiment}",
    eprint = "1307.6346",
    archivePrefix = "arXiv",
    primaryClass = "hep-ex",
    doi = "10.1007/JHEP02(2014)057",
    journal = "JHEP",
    volume = "02",
    pages = "057",
    year = "2014"
}

@inproceedings{Key4hep:2023nmr,
    author = "Sailer, Andre and others",
    collaboration = "Key4hep",
    title = "{The Key4hep software stack: Beyond Future Higgs factories}",
    eprint = "2312.08151",
    archivePrefix = "arXiv",
    primaryClass = "hep-ex",
    month = "December",
    year = "2023"
}

@article{CMS:2024onh,
    author = "Hayrapetyan, Aram and others",
    collaboration = "CMS",
    title = "{The CMS Statistical Analysis and Combination Tool: Combine}",
    eprint = "2404.06614",
    archivePrefix = "arXiv",
    primaryClass = "physics.data-an",
    reportNumber = "CMS-CAT-23-001, CERN-EP-2024-078",
    doi = "10.1007/s41781-024-00121-4",
    journal = "Comput. Softw. Big Sci.",
    volume = "8",
    number = "1",
    pages = "19",
    year = "2024"
}

@Article{HIGG-2021-23,
    author         = "{ATLAS Collaboration}",
    title          = "{A detailed map of Higgs boson interactions by the ATLAS experiment ten years after the discovery}",
    journal        = "Nature",
    volume         = "607",
    year           = "2022",
    pages          = "52--59",
    doi            = "10.1038/s41586-022-04893-w",
    reportNumber   = "CERN-EP-2022-057",
    eprint         = "2207.00092",
    archivePrefix  = "arXiv",
    primaryClass   = "hep-ex"
}

@Article{CMS-HIG-22-001,
    author         = "{CMS Collaboration}",
    title          = "{A portrait of the Higgs boson by the CMS experiment ten years after the discovery}",
    journal        = "Nature",
    volume         = "607",
    year           = "2022",
    pages          = "60--68",
    doi            = "10.1038/s41586-022-04892-x",
    reportNumber   = "CERN-EP-2022-039",
    eprint         = "2207.00043",
    archivePrefix  = "arXiv",
    primaryClass   = "hep-ex"
}

@misc{FCC:2025uan,
    author = "Benedikt, M. and others",
    collaboration = "FCC",
    title = "{Future Circular Collider Feasibility Study Report: Volume 2, Accelerators, Technical Infrastructure and Safety}",
    eprint = "2505.00274",
    archivePrefix = "arXiv",
    primaryClass = "physics.acc-ph",
    reportNumber = "CERN-FCC-ACC-2025-0004, FERMILAB-PUB-25-0909-PPD-TD",
    doi = "10.1140/epjs/s11734-025-01967-4",
    month = "April",
    year = "2025"
}

@misc{FCC:2025jtd,
    author = "Benedikt, M. and others",
    collaboration = "FCC",
    title = "{Future Circular Collider Feasibility Study Report: Volume 3, Civil Engineering, Implementation and Sustainability}",
    eprint = "2505.00273",
    archivePrefix = "arXiv",
    primaryClass = "physics.acc-ph",
    reportNumber = "CERN-FCC-ACC-2025-0003, FERMILAB-PUB-25-0911-PPD-TD",
    doi = "10.1140/epjs/s11734-025-01958-5",
    journal = "Eur. Phys. J. ST",
    volume = "234",
    number = "17",
    pages = "5113--5383",
    year = "2025"
}

@misc{FCC_EE_IDEA_Winter2023,
    author       = "{FCC Collaboration}",
    title        = "{IDEA---FCC-ee Winter 2023 Monte-Carlo Production}",
    year         = "2023",
    howpublished = "\url{https://fcc-physics-events.web.cern.ch/fcc-ee/rec/winter2023/IDEA}",
    note         = "Accessed December 14, 2025"
}

@misc{HEP-FCC_FCC-config,
    author       = "{FCC Collaboration}",
    title        = "{FCC-config: FCC configuration files}",
    year         = "2025",
    howpublished = "\url{https://github.com/HEP-FCC/FCC-config}",
    note         = "GitHub repository, accessed December 14, 2025"
}

@software{helsens_2025_15528870,
    author       = "Helsens, Clement and Perez, Emmanuel and Selvaggi, Michele and Volkl, Valentin and Forthomme, Laurent and Munch Torndal, Julie",
    title        = "{HEP-FCC/FCCAnalyses: v0.11.0}",
    month        = "May",
    year         = "2025",
    publisher    = "Zenodo",
    version      = "v0.11.0",
    doi          = "10.5281/zenodo.15528870",
    url          = "https://doi.org/10.5281/zenodo.15528870"
}

@inproceedings{Chen:2016:XGBoost,
    author    = "Chen, Tianqi and Guestrin, Carlos",
    title     = "{XGBoost: A Scalable Tree Boosting System}",
    booktitle = "{Proceedings of the 22nd ACM SIGKDD International Conference on Knowledge Discovery and Data Mining}",
    year      = "2016",
    pages     = "785--794",
    publisher = "ACM",
    doi       = "10.1145/2939672.2939785",
    eprint    = "1603.02754",
    archivePrefix = "arXiv"
}

@article{Brun:1997pa,
    author  = "Brun, Ren{\'e} and Rademakers, Fons",
    title   = "{ROOT --- An Object Oriented Data Analysis Framework}",
    journal = "Nucl. Instrum. Meth. A",
    volume  = "389",
    year    = "1997",
    pages   = "81--86",
    doi     = "10.1016/S0168-9002(97)00048-X"
}

\end{document}